\documentclass[
 preprint,
 amsmath,amssymb,
 aps,
 pof,
 nopreprintnumbers,
 showkeys,
]{revtex4-2}

\usepackage{graphicx}
\usepackage{dcolumn}
\usepackage{bm}
\usepackage{caption}
\usepackage{subcaption}
\newcommand{\volume}{\rotatebox[origin=c]{180}{\ensuremath{A}}}
\usepackage[normalem]{ulem}
\usepackage{xcolor}
\usepackage[noprefix]{nomencl}
\usepackage{url}
\usepackage{ulem}
\usepackage{natbib}
\usepackage[colorlinks,allcolors=blue]{hyperref}
\begin{document}

\preprint{APS/123-QED}

\title{The effect of droplet deformation and internal circulation on drag coefficient}

% \author{Yushu Lin}
% \author{John Palmore Jr}
% \affiliation{
%  Department of Mechanical Engineering, Virginia Tech\\
% }

\author{Yushu Lin}
 \email{linysh1997@vt.edu}
\author{John Palmore Jr.}
 \email{palmore@vt.edu}
\affiliation{
  Department of Mechanical Engineering\\
  Virginia Tech, Blacksburg, VA, USA
}

\date{\today}

\begin{abstract}
The current study uses numerical approaches to investigate the eﬀect of droplet deformation and internal circulation on droplet dynamics. Although droplet drag is a classical area of study, there are still theoretical gaps in understanding the motion of large droplets. In applications like spray combustion, droplets of various sizes are generated and move with the flow. Large droplets tend to deform in the flow, and have complex interactions with the flow because of this deformation. To better model spray, the physical understanding of droplets need to be improved. Under spray conditions, droplets are subjected to a high temperature and pressure environment, and the coupling between liquid and gas is enhanced. Therefore, the deformation and internal circulation will affect droplet drag coefficient more significantly than in atmospheric conditions. To study the mechanism on how droplet shape and internal circulation influence droplet dynamics, we have used direct numerical simulation (DNS) to simulate a droplet falling at its terminal velocity in high pressure air. An in-house code developed for interface-capturing DNS of multiphase ﬂows is employed for the simulation. The drag coefficient is calculated, and the results are consistent with existing literature for slightly deformed droplets. The results show that the drag coefficient is directly related to the droplet deformation and droplet internal circulation. The paper also develops an analytical theory to account for the effect of Weber number and fluid properties on droplet deformation.
\end{abstract}


\maketitle
\makenomenclature

\nomenclature{\(a\)}{Droplet center of mass acceleration}
\nomenclature{\(C_d\)}{Drag coefficient}
\nomenclature{\(D\)}{Droplet diameter}
\nomenclature{\(g\)}{gravity}
\nomenclature{\(P\)}{Gas pressure}
\nomenclature{\(P^*\)}{$P$ normalized by atmospheric pressure}
\nomenclature{\(t\)}{Time}   
\nomenclature{\(t_{cap}\)}{$P$ normalized by atmospheric pressure}
\nomenclature{\(t^*\)}{Time normalized by $t_{cap}$}
\nomenclature{\(U_{in}\)}{Inlet velocity of gas}
\nomenclature{\(U_d\)}{Droplet center of mass velocity}
\nomenclature{\(X_d\)}{Droplet center of mass displacement}

\section{Introduction}
Many problems in science and engineering involve the formation and motion of droplets. Common examples include spray painting, sneezing and disease prevention, fire suppression, and spray combustion. In these problems, accurately predicting how droplets move is important to predicting the efficacy of the engineering system. This work is particularly interested in predicting the motion of large droplets in these problems, as there are theoretical gaps into the behavior of such droplets. Hence it is a topic worth further investigation.

\begin{figure}[ht]
    \centering
    \includegraphics[width=0.6\textwidth]{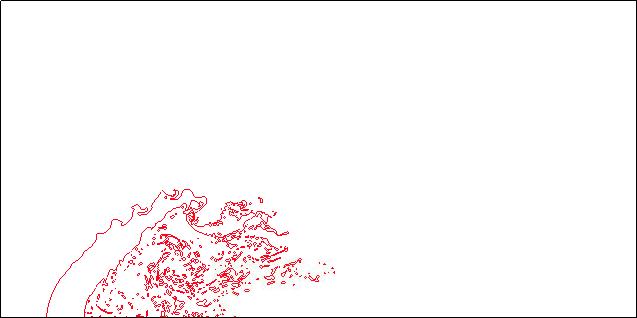}
    \caption{2D jet in cross-flow simulation using the parameters of \cite{herrmann_detailed_2010}}
    \label{jet}
\end{figure}

The research motivation for this work is the phenomenon of spray combustion in aviation gas turbine engines. Spray combustion consists a series of complex physical processes, including jet atomization (also called primary atomization), droplet breakup (also called secondary atomization), evaporation, droplet interaction, and combustion. In Figure.\ref{jet}, a 2D simulation is performed to show liquid jet atomizaiton \cite{palmore_interface-capturing_2022}. A liquid jet is injected into the crossflow, and droplets are produced from fragmentation of the jet. Although it is technically feasible to perform simulations of high-fidelity simulations of the type used in Fig.\ref{jet}, such simulations are largely restricted to use in specialized codes associated to academic and government research labs \cite{herrmann_detailed_2010,palmore_interface-capturing_2022,wen_atomization_2020}. Simulating droplet motion with these codes is too computationally expensive in practice.

An alternative approach looks at reduced order representations of droplets as Lagrangian particles. The common starting assumption for the Lagrangian model is that droplets are perfectly spherical and have no internal flow. This is accurate for the smallest spray droplets, however, in spray, droplets come in a range of sizes. The largest ones are large enough to see significant deformation which can fundamentally affect their behavior including drag \cite{loth_quasi-steady_2008} and evaporation rate \cite{palmore_vaporization_2022,setiya_draft}. Therefore, we need to improve the physical understanding of droplets to better predict the dynamics of droplets represented by Lagrangian particles. More specifically, this work performs a study on how the droplet drag coefficient is dependent on relevant parameters. However, the work does not do a wide parameter sweep of all scenarios possible. Instead, it focuses on understanding the physical mechanisms that govern droplet drag, and looks at several extreme conditions which succinctly demonstrate these principles.

The study of droplet drag coefficient has received attention from researchers for decades. The simplest approximation is to use the standard drag curve of rigid spheres as derived by Stokes. Several empirical correlations have been posed by multiplying a correction factor to Stokes' law $C_d=24/Re$. A review on these correlations can be found in \cite{goossens_review_2019}. For viscous liquid spheres, the droplet internal flow was assumed to be Hill's spherical vortex in \cite{harper_motion_1968}, while the gaseous flow was a potential flow. Based on these approximations, by integrating the surface stress, an analytical drag coefficient correlation of the first order approximation was derived depending on Reynolds number $Re$ and dynamic viscosity ratio $\mu^*=\mu_l/\mu_g$ in \cite{harper_motion_1968}. The subscript $l$ and $g$ represent liquid and gas respectively. For $Re$ up to 200 with arbitrary $\mu^*$, $C_d$ correlation for viscous spherical droplets was found from numerical simulations by the work in \cite{rivkind_flow_1976}, where the flow was considered steady and axisymmetric, and the Navier-Stokes equations were solved by solving the stream function and vorticity equation in a spherical coordinate. In \cite{feng_drag_2001}, a numerical method was developed by introducing a two-layer concept to capture the very thin boundary layer at the liquid-gas interface, and a drag coefficient for viscous spherical droplets was well established for intermediate $\mu^*=\mu_l/\mu_g$. In \cite{loth_quasi-steady_2008}, drag correlation for deformed droplets were found by examining and fitting experimental results in \cite{reinhart_verhalten_1964}. For drag coefficient correlations of deformed droplets, one can refer to \cite{haywood_numerical_1994}. In the work of \cite{haywood_numerical_1994}, a finite-volume method was used in a non-orthogonal adaptive grid system, and the energy equation was solved as well for droplet evaporation. To investigate the effect of droplet internal circulation, it was revealed in \cite{law_theory_1977} by scaling analysis that the multiplication of density ratio $\rho^*=\rho_l/\rho_g$ and dynamic viscosity ratio $\Lambda=\sqrt{\rho^*\mu^*}$ characterizes the coupling between the liquid and gas phases. However, \cite{law_theory_1977} refers to a spherical droplet. \citet{helenbrook_quasi-steady_2002} and \citet{feng_deformable_2010} investigated both the effect of deformation and internal circulation of droplets. An arbitrary-Lagrangian-Eulerian mesh movement scheme with unstructured mesh was used in \cite{helenbrook_quasi-steady_2002} to resolve the position of phase interface. In \cite{feng_deformable_2010}, it was found that given $Re$ and Weber number $We$, droplet drag coefficient was dependent only on $\rho^*/\left(\mu^*\right)^2$, which was equivalent to Ohnesorge number $Oh$ in such case. The Navier-Stokes equations were solved by Galerkin finite-element method in a cylindrical coordinate. The aforementioned works focus on steady droplets. For a spherical droplet that is accelerating or decelerating in the gas, in \cite{temkin_droplet_1980}
and \cite{temkin_droplet_1982}, a correlation to $C_d$ was found based on conical-driver shock tube experiment results, and it was dependent on a non-dimensional relative-acceleration parameter. For deformed transient droplets, the $C_d$ correlation can be found in \cite{qu_numerical_2016}, where the ANSYS Fluent was used to solve the Navier-Stokes equations.

For gas turbine combustion, classical droplet models are incomplete, because in the high temperature and pressure environment, droplets are usually highly deformed, and the coupling between gas and liquid phases is enhanced. These phenomena result in different behaviors between droplets in spray and the droplet theory, consequently, a detailed study on both the effect of droplet shape and internal circulation effect will be needed. To study their effect on the droplet drag coefficient, we utilize an in-house code developed by our group \cite{palmore_volume_2019} for interface-capturing DNS of vaporization multiphase flows. The code uses volume-of-fluid method (VOF) to determine the location of phase interface, and solves Navier-Stokes equations in the whole domain in Cartesian coordinate. Transient droplet motion can be calculated by the code directly. A result for this work is that a more accurate drag coefficient calculation is found by correcting the droplet frontal area estimation which agrees with some previous literature. The novelty of this study is that the 3D code simulates droplet shape deformation from first principle without any assumption, and it also includes the effect of pressurized gas on the internal circulation of droplets. More details about numerical implementations can be found in Sec.\ref{method}.

\section{Problem formulation}
\begin{figure}[ht]
    \centering
    \includegraphics[width=0.4\textwidth]{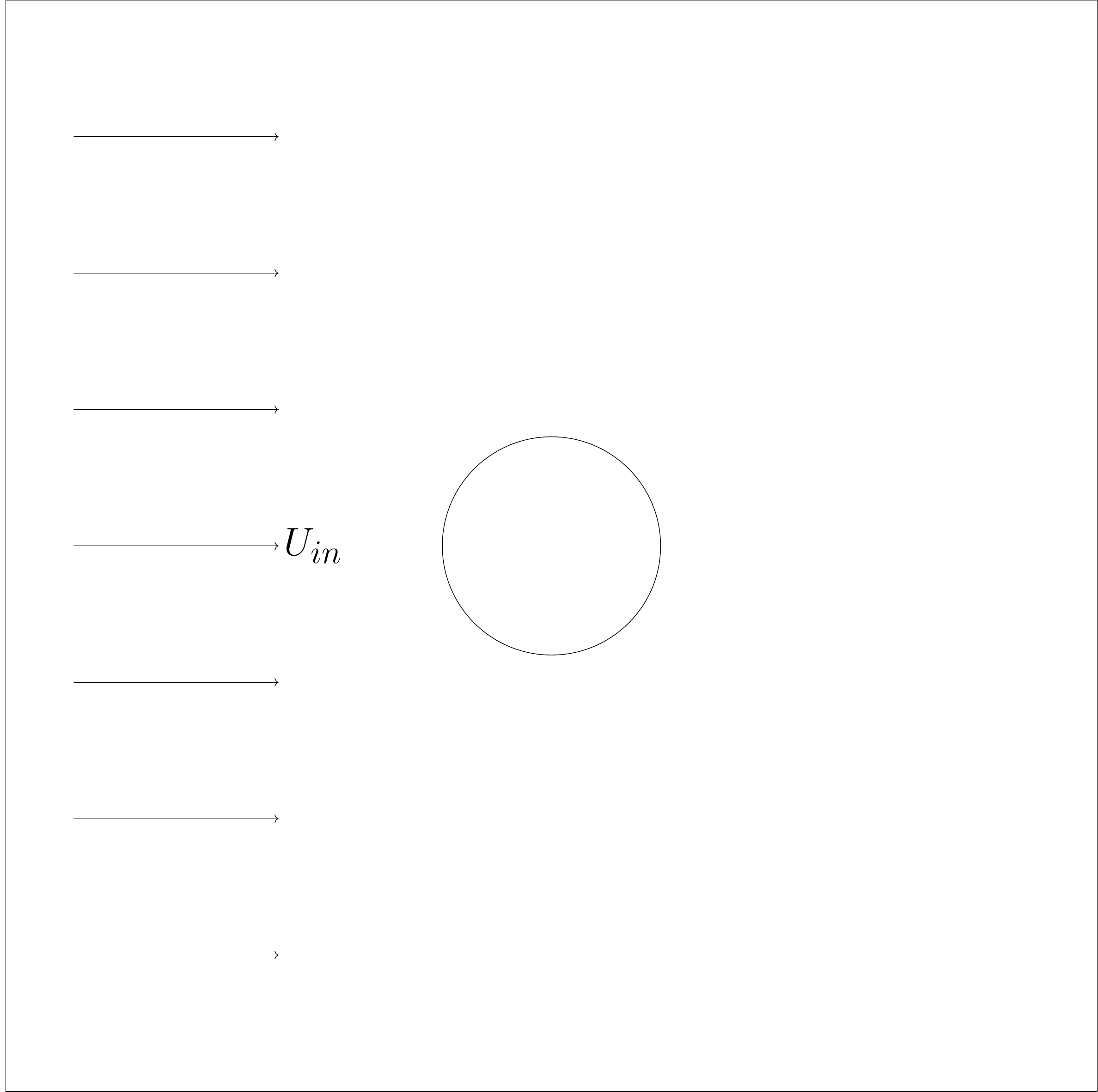}
    \caption{2D slice of the computation domain}
    \label{config}
\end{figure}

We simulate droplet deformation in a uniform convective flow in 3D. An initially spherical n-decane droplet with diameter $D$ is centered at a cubic computation domain with size $\left(8D\right)^3$. Dry air flow enters the domain uniformly in a speed of $U_{in}$ from inlet boundary, and leaves the domain freely on the exit boundary, see Fig.\ref{config}. Periodic boundary conditions are applied to other sides of the domain boundaries. We choose the droplet center of mass as the frame of reference, such that the droplet will be stationary at the center in the domain. To achieve this, an artificial gravity will be used to balance the drag force, and the droplet will reach its terminal velocity. The artificial gravity will be discussed with more details in Sec.\ref{gravity}.

\subsection{Controlling parameters}
Based on the work in \cite{guildenbecher_secondary_2009} and \cite{law_theory_1977}, the following non-dimensional groups are determined to be the controlling parameters in our problem:
\begin{equation}
Re=\frac{\rho_gU_{in}D}{\mu_g},
\end{equation}
\begin{equation}
We=\frac{\rho_gU_{in}^2D}{\sigma},
\end{equation}
\begin{equation}
\Lambda=\sqrt{\frac{\rho_l\mu_l}{\rho_g\mu_g}}.
\end{equation}

The Reynolds number compares the inertial force and viscous force of gas flowing past the droplet. The Weber number compares the inertia of the gas and the surface tension of liquid, indicating how well a droplet can keep itself spherical. $\Lambda$ reveals the strength of internal motion of droplet compared to the freestream gas flow. In a real flow, $\Lambda$ will vary due to the changes in temperature and pressure of the liquid and gas. Comparing atmospheric conditions to those in a gas turbine engine, the dominant effect in $\Lambda$ is due to gas density change accompanying the high gas pressurization. With a fixed Reynolds number, we can control droplet shape by changing Weber number or control internal flow by changing liquid-to-gas density.

\subsection{Configurations}\label{setup}
We perform various numerical simulations on droplet falling at its terminal velocity at $Re=70$. To study its deformation and internal circulation, we must ensure that the droplet does not breakup. In all our cases, Ohnesorge number $Oh$ is less then $0.1$, so that droplet will not breakup for $We<We_{\text{critical}}\approx12$ \cite{suryaprakash_secondary_2019}. The simulation parameters for the setup are listed in Table.\ref{parameter}. The dynamic viscosity is $2.47425\times10^{-5}\ \text{kg}/\left(\text{m}\cdot\text{s}\right)$ for gas at liquid-gas interface temperature, which is the boiling temperature of n-decane at around 447.3K \cite{noauthor_air_2003}, and $2.0241\times10^{-4}\ \text{kg}/(\text{m}\cdot\text{s})$ \cite{pj_nist} for liquid. The density of liquid is $603.87\ \text{kg}/\text{m}^3$ \cite{pj_nist}, and the density of gas is determined by the density ratio. Surface tension is set to be $0.01024\ \text{kg}/\text{s}^2$ \cite{pj_nist}. The properties of the liquid depends weakly on pressure, so we treat liquid properties as constants. The droplet diameter is determined to achieve parameters set in Table.\ref{parameter}, and the domain size is chosen in proportion to the diameter. 

\begin{table}[!ht]
\centering
\begin{tabular}{||c|c|c|c||} 
 \hline
 Case & $\rho^*$ & $P^*=P/P_{atm}$ & $We$ \\ [0.4ex] 
 \hline\hline
a1$\sim$a4 & 20 & 38.2 & 1,\ 3,\ 6,\ 9 \\ 
b1$\sim$b4 & 20,\ 40,\ 60,\ 160,\ 765 & 38.2,\ 19.1,\ 9.56,\ 4.78,\ 1.0 & 6 \\ 
 [1ex] 
 \hline
\end{tabular}
\caption{Properties of the gas and liquid}
\label{parameter}
\end{table}

\section{Numerical methods} \label{method}
We employ an in-house code called NGA \cite{desjardins_high_2008} to simulate the droplet falling at its terminal velocity. NGA is developed for solving low-Mach number turbulent flows, and is further developed for interface-capturing multiphase flows by \citet{palmore_volume_2019}. This section will give a brief overview of some of the algorithms.
\subsection{Governing equations}
The governing equations for conservation of momentum in both liquid and gas phases are:
\begin{equation}
        \frac{\partial\left(\rho\bm{u}\right)}{\partial t}+\nabla\cdot\left(\rho\bm{u}\otimes\bm{u}\right)=-\nabla p+\nabla\cdot{\mu\bm{S}}+\bm{f},\label{momentum}
\end{equation}
where $\bm{S}=\nabla\bm{u}+\nabla\bm{u}^{\top}-\frac{2}{3}\nabla\cdot\bm{u}$, $\bm{f}$ is the external body force used for stabilizing droplet at the center. The conservation of mass is guaranteed by solving the pressure term from pressure Poisson equations. The pressure Poisson equation is solved by using the Ghost Fluid Method(GFM) \cite{liu_boundary_2000}. In Eq.\ref{momentum} $\rho$ and $\mu$ are effective density and effective viscosity introduced in Sec.\ref{interface} for solving governing equations in one-phase approach.

\subsection{Interface capturing method}\label{interface}
To solve the liquid-gas interface, the VOF method is employed. In VOF method, volume fraction $\phi$ is defined in each cell as the volume fraction occupied by the liquid. For cells entirely within the liquid phase, $\phi=1$, and for cells within the gas phase, $\phi=0$. When the phase interface crosses a cell, the volume fraction will be within the range of $\left(0,1\right)$, and the effective density and viscosity in these cells can be defined as: 
\begin{equation}
    \begin{aligned} 
        \rho&=\rho_l\phi+\rho_g\left(1-\phi\right),\\
        \mu&=\mu_l\phi+\mu_g\left(1-\phi\right).
    \end{aligned}
\end{equation}
The effective density and viscosity are used to solve governing equations of the gas and liquid in a one-phase approach \cite{palmore_volume_2019}. In this way, the velocity field of gas and liquid are treated as a union, and the governing equations can be solved only once for one unified velocity field. The evolution of the volume fraction scalar field is governed by the following advection equation with the velocity field to be the gas-liquid union velocity field \cite{owkes_computational_2014}:
\begin{equation}
    \frac{\partial\phi}{\partial t}+\bm{u}\cdot\nabla\phi=0.
\end{equation}

\subsection{Jump conditions across interface}
To ensure the conservation of mass and momentum at the phase interface, several matching conditions at phase interface should be satisfied. The current simulations are for non-evaporating droplets, so the only jump condition is the pressure jump due to surface tension: 
\begin{equation}
    P_g-P_l=-\sigma\kappa.
\end{equation}
where $\sigma$ is liquid surface tension, $\kappa$ is the curvature of the droplet surface, and it is defined so that $\kappa>0$ for convexly shaped liquid regions.

\subsection{Artificial gravity}\label{gravity}
The droplet in the computational domain will move due to the drag force. In the desire of studying a stationary droplet, \citet{palmore_validating_2018} devised a method that mimics the flow over a falling droplet at terminal velocity. In this method, the gravity is fixed and the terminal velocity is converged. However, due to the uncertainty of the drag coefficient caused by droplet deformation, the velocity reached will still be lower than the terminal velocity. To ensure constant terminal velocity was reached, \citet{setiya_method_2020,lin_numerical_2022} developed a gravity update scheme to balance the changing drag force. Since the drag force is unknown explicitly, their method is based on a feedback control loop:
\begin{equation}
    g^{n+1}=g^{n}+k_U U_d+k_X X_d,
    \label{setiya_gravity}
\end{equation}
where $U_d$ is the droplet velocity, $X_d$ is the droplet center of mass position in x-direction, $\alpha$ is an arbitrary weighting constant, and $\tau_c$ is the capillary time defined as $\tau_c=\sqrt{\frac{\rho_l+\rho_g}{\sigma}}\left(\frac{D}{2\pi}\right)^{\frac{3}{2}}$. $k_U$ and $k_X$ are gains of $U_d$ and $X_d$, respectively. This approach has higher robustness on inflow boundary conditions compared to \cite{palmore_validating_2018}. For the current setting, we set $k_U=1/\left(2\tau_c\right)$ and $K_X=0$. 

\subsection{Quantification of internal circulation}\label{circulation}
\begin{figure}[ht]
    \centering
    \includegraphics[width=0.7\textwidth]{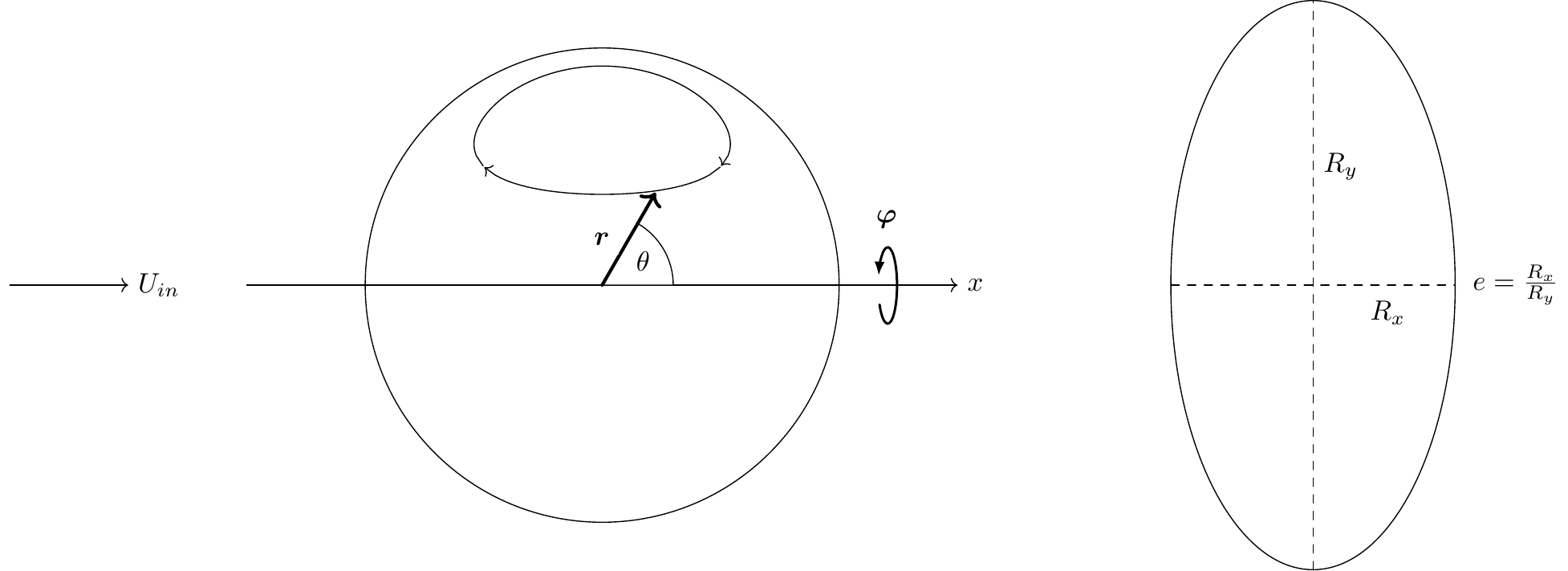}
    \caption{Coordinate system of spherical and deformed droplets}
    \label{coordinate}
\end{figure}

Because the increase of pressure in gas turbine can be significant, the gas density will increase proportionally through ideal gas law. On the contrary, the liquid density remains almost constant regardless of the pressure change. Hence, the gas density change represents the physical process of the pressure change. In addition, studies such as \cite{law_theory_1977,feng_deformable_2010,lin_numerical_2022} have revealed that internal circulation is dependent on density ratio. Therefore, we choose the gas density to be the controlling parameter for changing the strength of internal circulation. To quantify the strength of internal circulation, several common physical quantities are considered, including the maximum liquid velocity, the maximum and mean vorticity of the droplet, and the droplet enstropy. Based on our previous work, maximum velocity and vorticity are 
not good choices due to their sensitivity to numerical errors, while volume-averaged variables are better in representing internal circulation strength \cite{lin_numerical_2022}. Thus, we only use volume-averaged variables in this study. The definition of vorticity $\bm{\omega}$ and enstrophy $\mathcal{E}$ is:
\begin{align}
    \bm{\omega} &= \nabla\times \bm{U}_{L},\\
    \mathcal{E} &= \frac{1}{2}\bm{\omega}^2,
\end{align}
where $U_L$ is the liquid phase velocity field. $\omega$ is vorticity, and $\mathcal{E}$ is enstrophy. In addition, a variable we have termed the Hill's constant is also used as a measure of internal circulation strength. The Hill's spherical vortex \cite{panton_incompressible_2013} is the simplest modelling to the droplet internal circulation, and the Hill's constant is a quantity derived from Hill's solution to represent the internal circulation strength. The vorticity magnitude of Hill's vortex is given by \citet{batchelor_introduction_2000}:
\begin{equation}
    \left|\bm{\omega}\right|=Ar_\perp,
\end{equation}
where $\bm{r}$ is a vector within the droplet from  droplet center, $\theta$ is the angle enclosed by $\bm{r}$ and the x-axis, $r_\perp=\left|\bm{r}\right|\sin\theta$ and $A$ is the Hill's constant representing the vortex strength. In our code, the local Hill's constant is calculated by using the following expression:
\begin{equation}
    A=\frac{\bm{\omega}\cdot\hat{\bm{{\varphi}}}}{r_\perp},
\end{equation}
where $\hat{\bm{\varphi}}$ is the unit vector normal to $\bm{r}$ in ${\bm{\varphi}}$ direction, see Fig.\ref{coordinate}.

When calculating the mean vorticity, enstrophy and Hill's constant, we will implement a volume integration over the droplet: 
\begin{equation}
    \bar{\psi}=\frac{1}{V} \iiint_V \psi \,d\volume,
\end{equation}
where $V$ is the volume of the droplet, $\psi$ is any variable that need to be volume-averaged, and $\bar{\psi}$ is the volume-averaged variable.

A limitation of numerical strategy for momentum solution is that liquid and gas velocity are solved on one field. For cells near the phase interface, large velocity difference might occur if gas velocity is included. To avoid including gas velocity in the calculation of circulation variables, the liquid phase velocity field $U_L$ is computed as a post processing step at each timestep from the combined velocity. This value is smoothly extrapolated into gas using the technique of \cite{aslam_partial_2004}, and the smoothed field is the one used for the computations of vorticty, enstrophy, and Hill's constant.

\subsection{Calculation of drag coefficient}
Assume the droplet is stationary in the domain, and the droplet motion is dominated in $x$-direction then drag coefficient $C_d$ can be computed by balancing the drag force and gravity:
\begin{equation}\label{drag_old}
    \frac{1}{2}C_d\rho_GU_{in}^2A_p=\left(\rho_L-\rho_G\right)gV,
\end{equation}
where $A_p$ is the projected droplet frontal area calculated by using effective radius $r_{\textrm{eff}}$. $r_{\textrm{eff}}$ is the radius of a spherical droplet which has the same volume with the deformed droplet. Since we are not working on evaporation in the current study, droplet volume and $r_{\textrm{eff}}$ will not change, thus $A_p$ is a constant.

Some further modifications can be made on the drag coefficient estimation. At first approximation, we can assume that the deformed droplet is spheroid \cite{loth_quasi-steady_2008}, and use aspect ratio $e$ of the spheroidal droplet to calculate $A_p$ more accurately. In our study, we define $e$ as the ratio of semi-axis length in $x$ and $y$ directions, i.e. $e=R_x/R_y$, see Fig.\ref{coordinate}. In the code, we estimate the semi-axis length by choosing the maximum of the summation of volume fraction at each line along the diction. Secondly, if the droplet is stationary at the center of the domain as expected, the terminal velocity will be equal to the inlet gaseous velocity. However, after experiencing the initial transient period, often the droplet will move very slowly at a nearly constant speed even with the gravity update scheme discussed in Sec.\ref{gravity}. This cannot be avoided since motion by a constant velocity satisfies the Navier-Stokes equations via its Galilean invariant property. To account for this, we can replace $U_{in}^2$ by $\left(U_{in}-U_d\right)^2$ to improve the accuracy of terminal velocity estimation. Here $U_d$ is the droplet average value of the x-component of $U_L$. Strictly, Eq.\ref{drag_old} requires the droplet to be non-accelerating. To quantify the effect of the small droplet acceleration, we also include the acceleration term in the calculation of droplet coefficient:
\begin{equation}\label{drag_new}
\begin{aligned}
    \frac{1}{2}C_d\rho_G&\left(U_{in}-U_{d}\right)^2A_p\\
    =&\left(\rho_L-\rho_G\right)gV+\rho_LaV,
\end{aligned}
\end{equation}

An advantage to this approach is that there could be other terms affecting droplet drag other than gravity. For example, in \cite{maxey_equation_1983} and \cite{crowe_particle-fluid_2011}, it is mentioned that the added mass term and history term can play a role in certain circumstances. These terms are usually negligible in steady-state problems, however, they may be important for the initial transient portion of the flow. Since the acceleration is calculated directly from droplet motion, the acceleration term automatically captures all effects that are not explicitly given in Eq.\ref{drag_old}. Therefore, the effect of added mass term and history term are implicitly included in the acceleration term.

To find an accurate calculation of acceleration $a$, we have tried three different ways to compute it. We label them as $a_1$, $a_2$ and $a_3$: $a_1=dU_d/dt$, $a_2=d^2X_d/dt^2$ and $a_3=d\left(U_d^2\right)/2dX_d$. $X_d$ is the centroid of the droplet. The three potential definitions were chosen to control numerical error of the approximation used to compute the acceleration. $a_1$ is the most straightforward definition of the acceleration. However, previous simulations have shown this may not be a perfect representation of droplet motion, because $U_d$ does not exactly represent the motion of the droplet centroid due to errors in the extrapolation process used to define $U_L$. $a_2$ computes the acceleration directly from the droplet position, but is a slightly more noisy value. For example, a pinned droplet that oscillates in place will demonstrate changes in $a_2$ due to slight asymmetries in the interface shape. $a_3$ is borrowed from 1D kinematics of particles. Since it combines $U_d$ and $X_d$, it has the potential to control for errors in either of the other definitions.
The discretized form of each is:
\begin{equation}
    a_1^n=\frac{U_d^{n+1}-U_d^{n-1}}{2\Delta t},
\end{equation}
\begin{equation}
    a_2^n=\frac{X_d^{n+1}-2X_d^n+X_d^{n-1}}{\Delta t^2},
\end{equation}
\begin{equation}
    a_3^n=\frac{\left(U_d^{n+1}\right)^2-\left(U_d^{n-1}\right)^2}{2\left(X_d^{n+1}-X_d^{n-1}\right)},
\end{equation}
where the superscript $n$ denotes for the $n$-th timestep.

\section{Results and discussions}
In this section, based on parameters explored in Sec.\ref{setup}, we will discuss the results of our simulations to see how liquid-to-gas density ratio and Weber number affect droplet deformation and drag coefficient.

\subsection{Grid convergence study}\label{grid}
\begin{figure}[ht]
    \begin{subfigure}{0.48\textwidth}
        \centering
        \includegraphics[width=\textwidth]{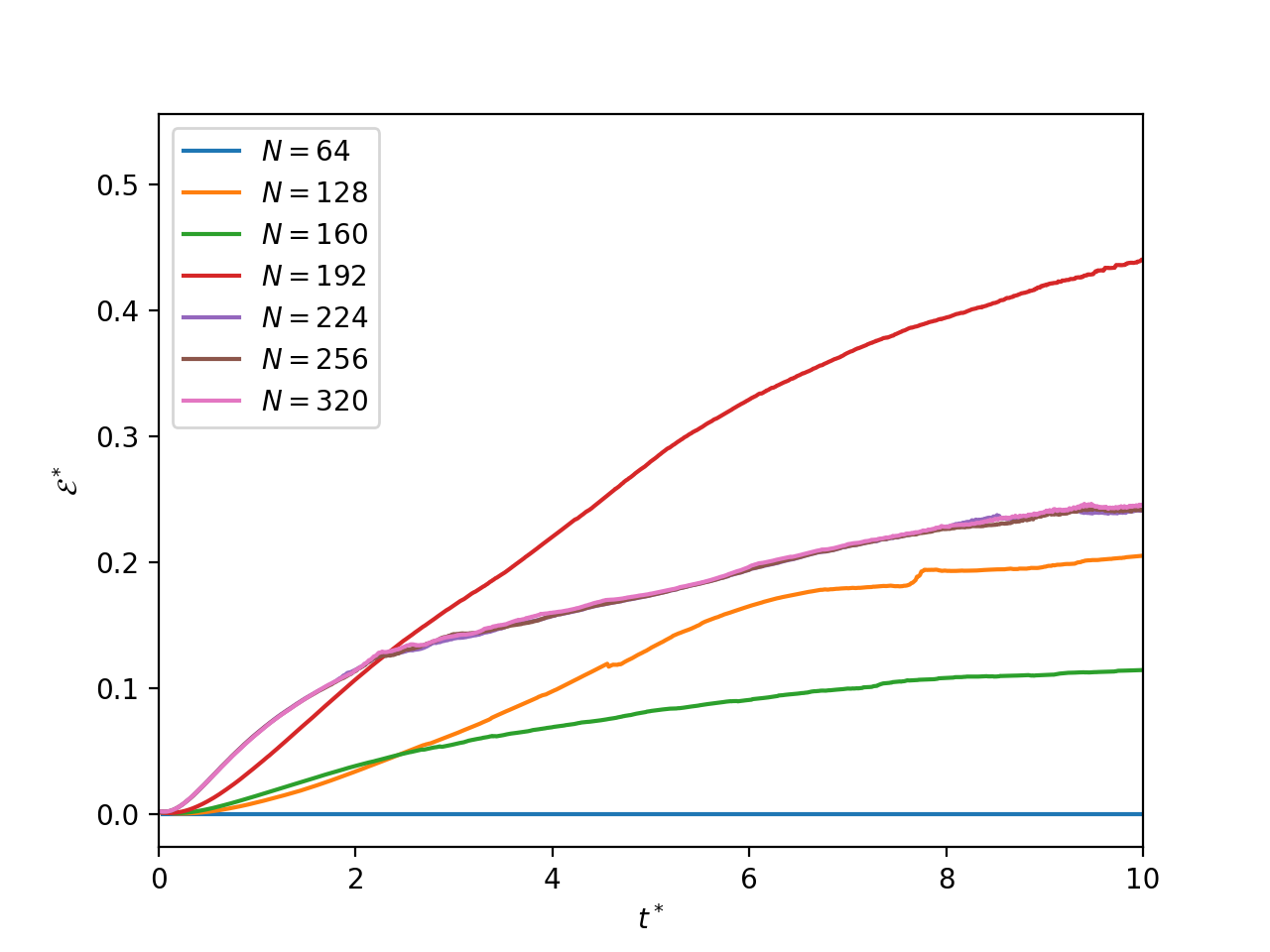}
        \caption{Nondimensionalized volume-averaged enstrophy (definition of nondimensionalization is in caption of Fig.\ref{circulation-t})}
        \label{grid-enstrophy}
    \end{subfigure}
    \begin{subfigure}{0.48\textwidth}
        \centering
        \includegraphics[width=\textwidth]{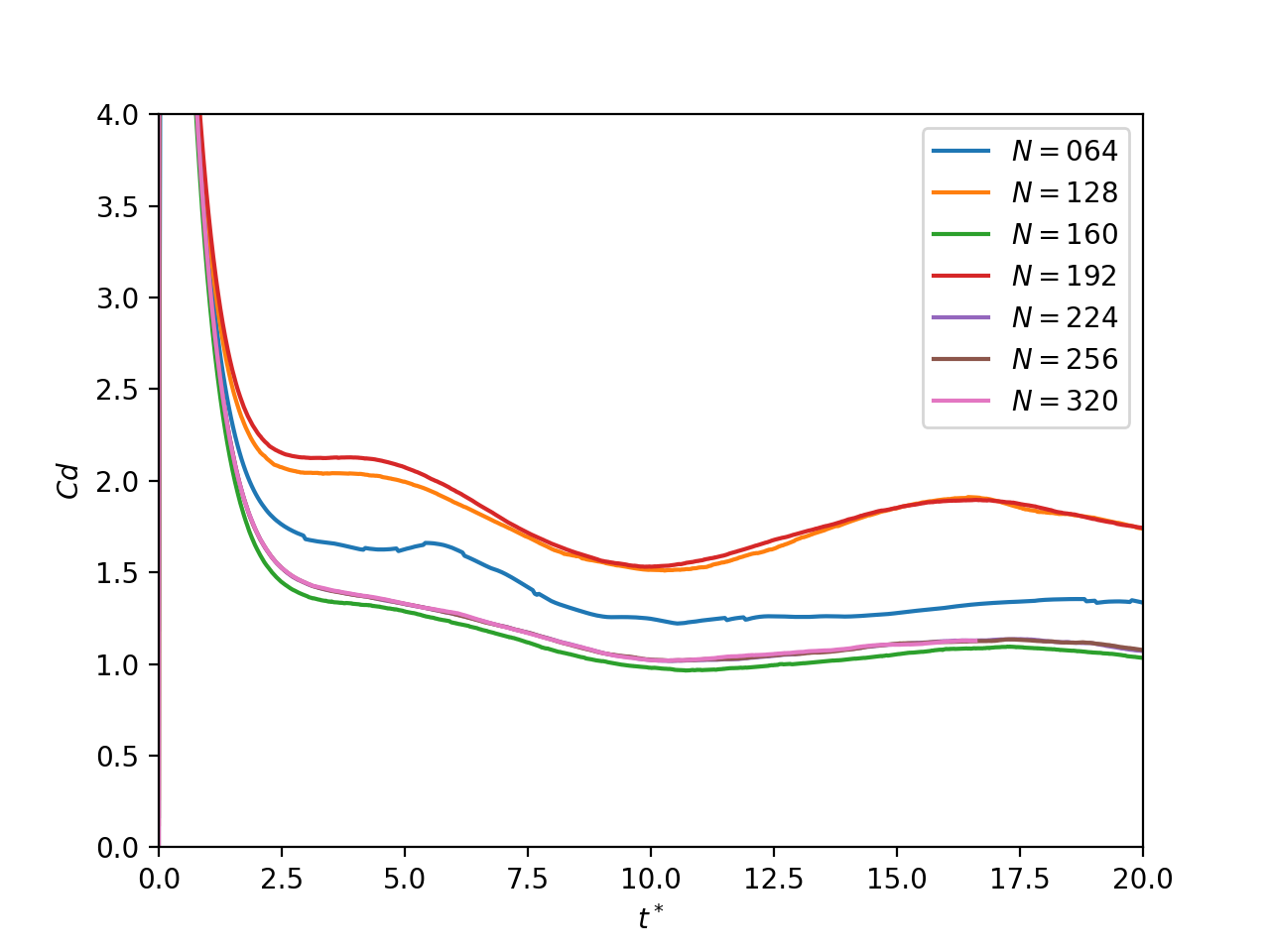}
        \caption{Drag coefficient calculated by Eq.\ref{drag_old} with $A_p$ estimated by $r_{\text{eff}}$\newline}
        \label{grid-steadCd}
    \end{subfigure}
    \caption{Enstrophy and drag coefficient on different grids}
    \label{grid_study}
\end{figure}

It will require a very fine mesh to resolve the liquid-gas interface and internal flow inside the droplet. Therefore, to eliminate the influence of grid resolution, we will first perform simulations with different grids to determine a suitable mesh resolution. The flow conditions selected for grid convergence study is $We=1$, $Re=70$ and $\rho^*=20$. In this case, since droplet is nearly spherical, droplet internal circulation will not be further affected by droplet deformation, so the mesh size will be the only factor that influences internal flow. We increase grid points from $N=64$ to $N=320$ on each dimension, and examine results of drag coefficient and enstrophy of droplet, see Fig.\ref{grid_study}. At very coarse meshes, the results vary significantly, however, the results finally converged at $N=224,256,320$, i.e. drag coefficient and enstrophy will not change anymore with increasing grid points for $N>224$. Therefore, we finally chose $N=256$ for our simulations.

\begin{figure}[ht]
    \begin{subfigure}{0.32\textwidth}
    \includegraphics[width=\textwidth]{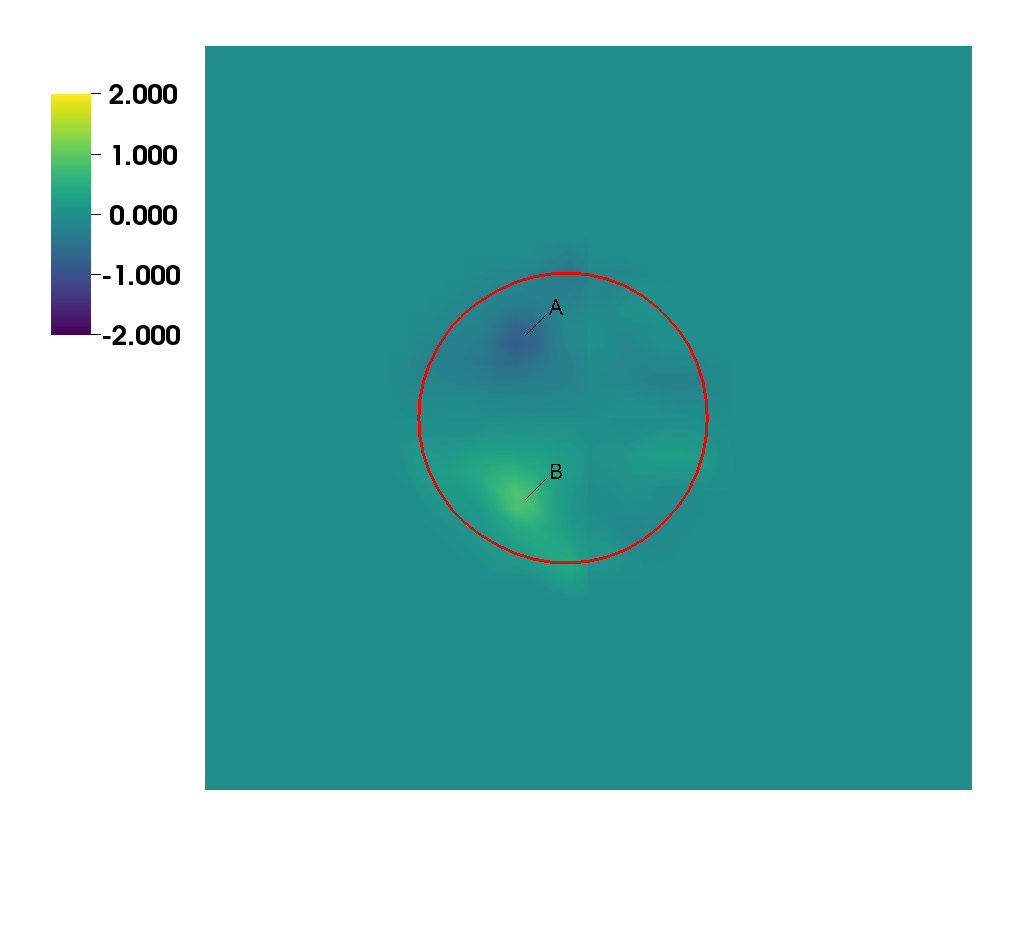}
        \caption{$N=128$}
        \label{N128}
    \end{subfigure}
    \begin{subfigure}{0.32\textwidth}
    \includegraphics[width=\textwidth]{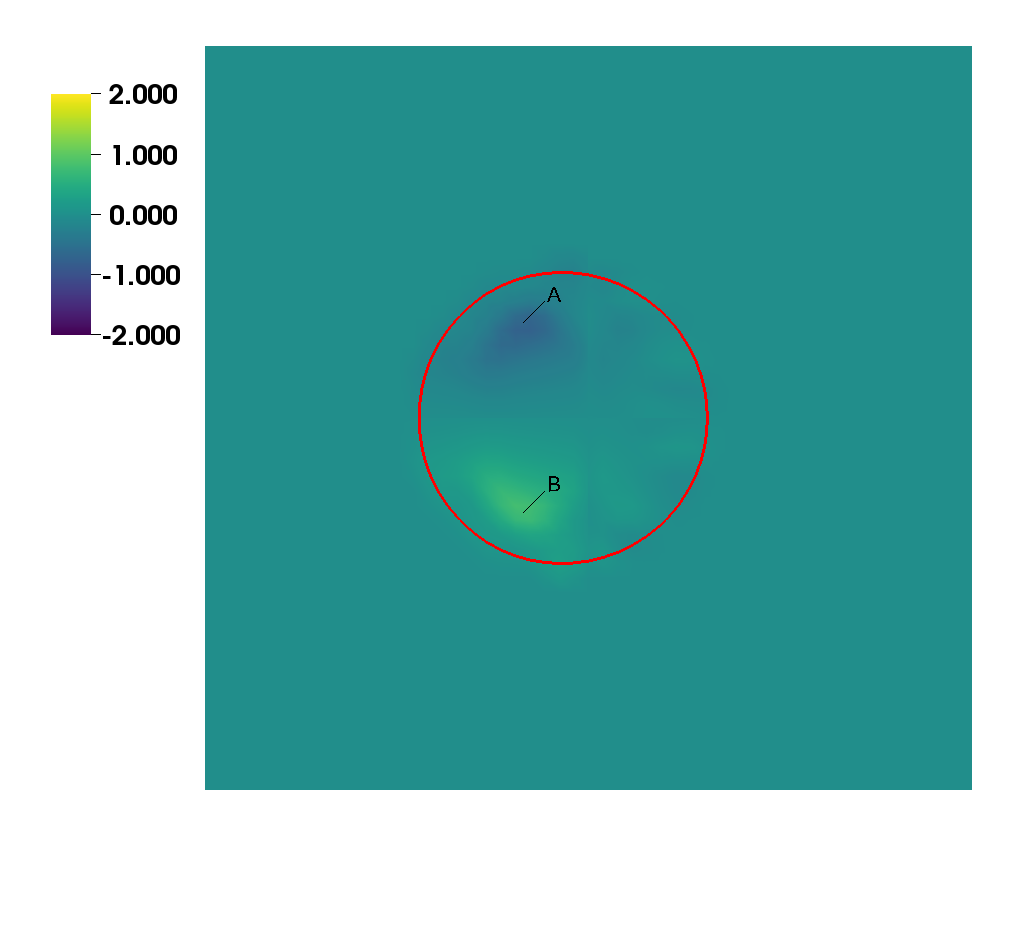}
        \caption{$N=160$}
        \label{N160}
    \end{subfigure}
    \begin{subfigure}{0.32\textwidth}
    \includegraphics[width=\textwidth]{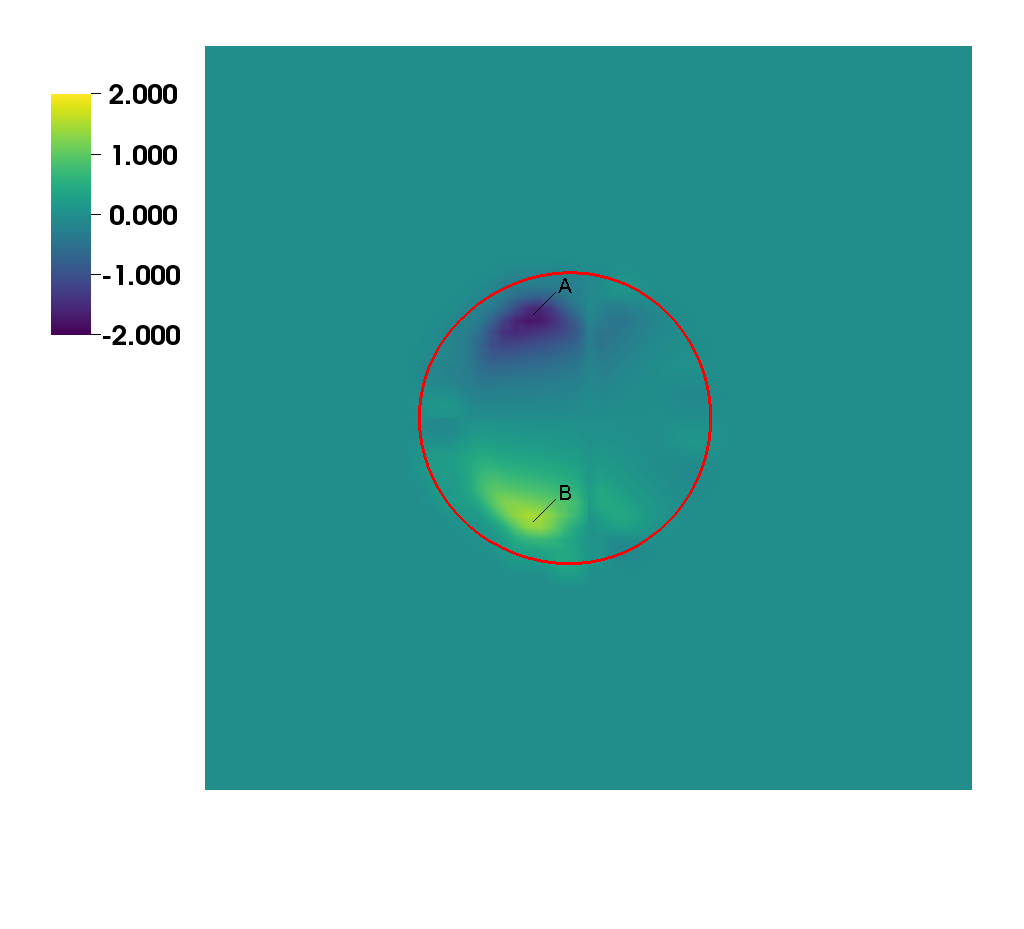}
        \caption{$N=192$}
        \label{N192}
    \end{subfigure}
    \begin{subfigure}{0.32\textwidth}
    \includegraphics[width=\textwidth]{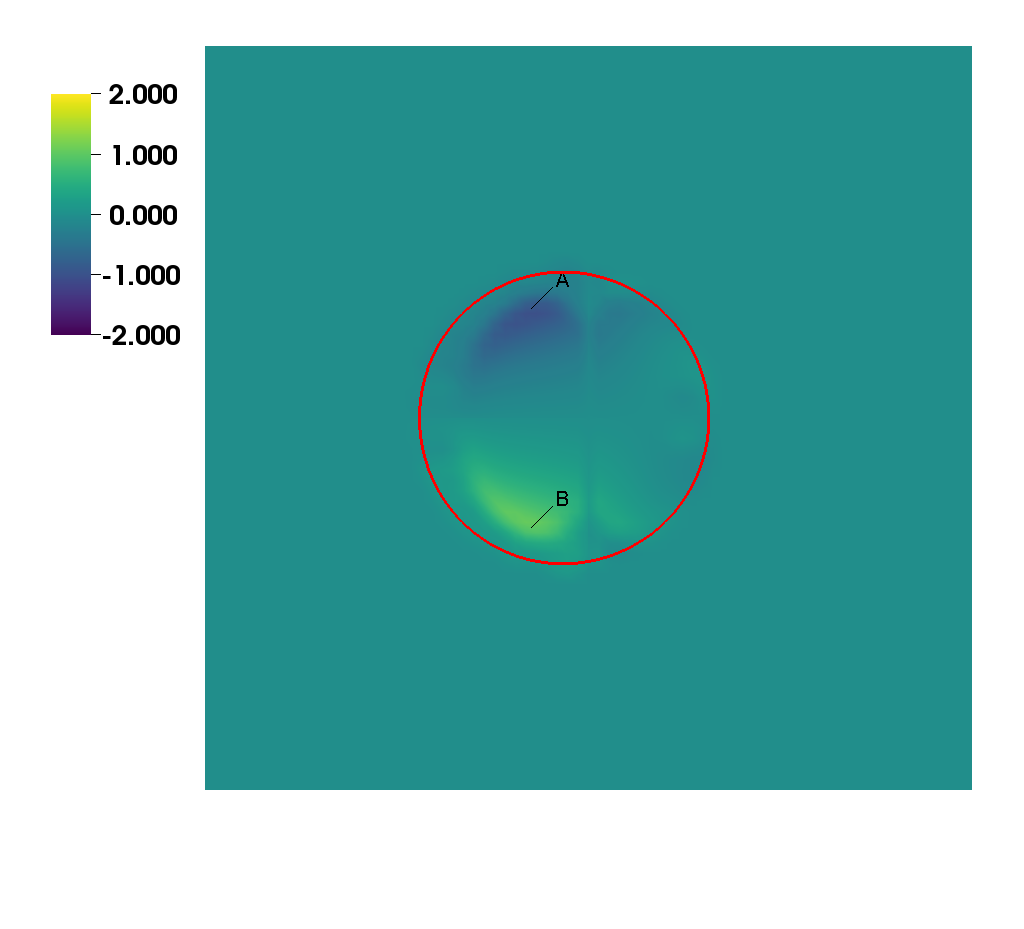}
        \caption{$N=224$}
        \label{N224}
    \end{subfigure}
    \begin{subfigure}{0.32\textwidth}
    \includegraphics[width=\textwidth]{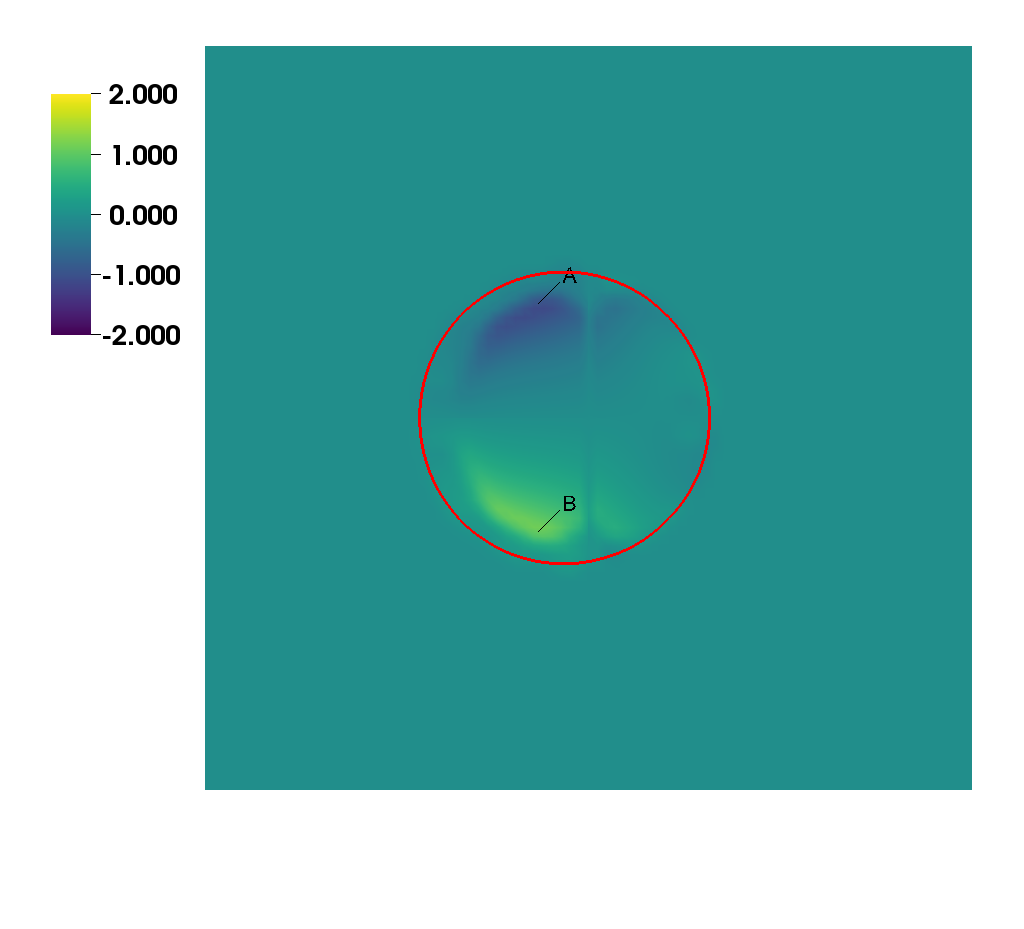}
        \caption{$N=256$}
        \label{N256}
    \end{subfigure}
    \begin{subfigure}{0.32\textwidth}
    \includegraphics[width=\textwidth]{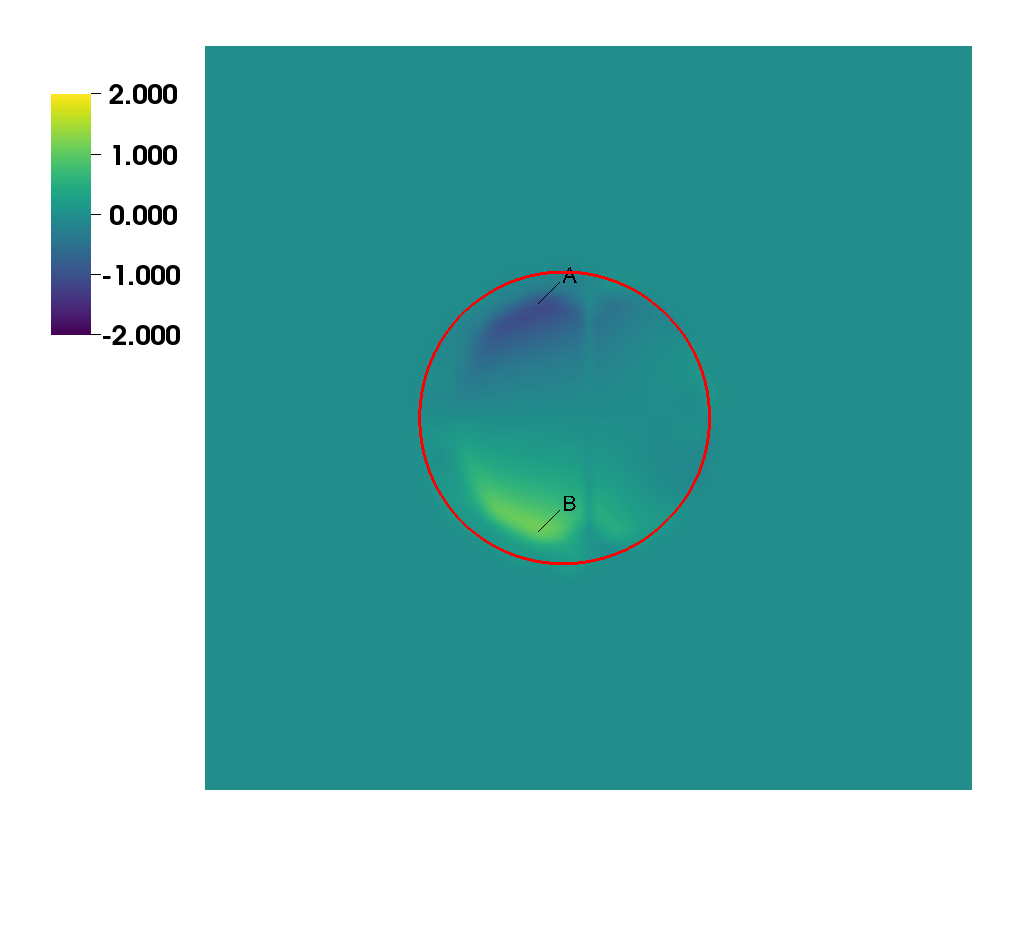}
        \caption{$N=320$}
        \label{N320}
    \end{subfigure}
    \caption{Pseudocolor of nondimensionalized magnitude of z-component vorticity; A and B denote for maximum value of vorticity (definition of nondimensionalization is in caption of Fig.\ref{circulation-t}), and red circle is liquid-gas interface}
    \label{grid_vortex}
\end{figure}

It is interesting to note that the results shown in Fig.\ref{grid_study} do not demonstrate a monotonic trend as mesh resolution increases. To explain this, we further examined the internal structure of droplet. As shown in Fig.\ref{grid_vortex}, the vorticity in z-direction are plotted.The complex interaction between the core vortex dynamics and the boundary layer dynamics contribute to the non-monotonic behavior of internal circulation strength and drag coefficients observed in Fig.\ref{grid_study}. As mesh resolution increases, the largest vorticity location moves closer to the boundary. It appears that for $N=128$, the largest vorticity location is very close to the center of the top and bottom semi-sphere of droplet. However, the core vortex region (deep blue and bright yellow parts) spreads from a very compact region in $N=128$ to an arc-like structure in $N=320$. As the high speed circulating fluid from the vortex moves closer to the surface, the liquid boundary layer at the liquid-gas interface must become thinner. This causes a competing action where as the vortex core is better resolved it moves, and this causes the boundary layer to be resolved more poorly. However, with sufficiently fine mesh resolution the vortex structure no longer changes, and the boundary layer can be resolved. It is interesting to note, that these dynamics also affect the shapes of the droplets. For $N=128$ and $N=192$, the curvature of interface on top-left and bottom-left parts are flatter than $N=160$. For $N=224,256,320$, their shapes do not alter too much. In the end, both the vortex structure and droplet shape of $N=256$ and $N=320$ are almost identical, which suggests using $N=256$.

\subsection{Changing $We$ at fixed $\rho^*$}\label{result-We}
\begin{figure}[!ht]
    \begin{subfigure}{0.4\textwidth}
    \includegraphics[width=\textwidth]{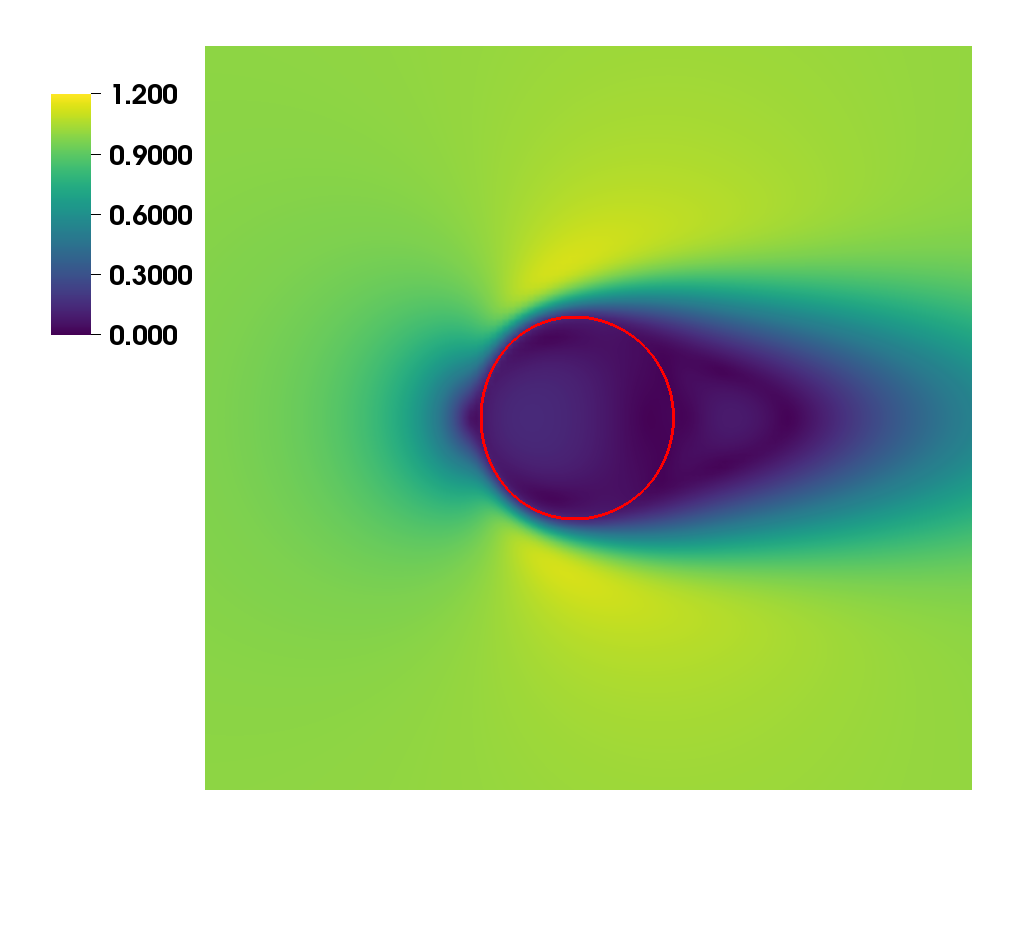}
        \caption{$We=1$}
        \label{drop-we1}
    \end{subfigure}
    \begin{subfigure}{0.4\textwidth}
    \includegraphics[width=\textwidth]{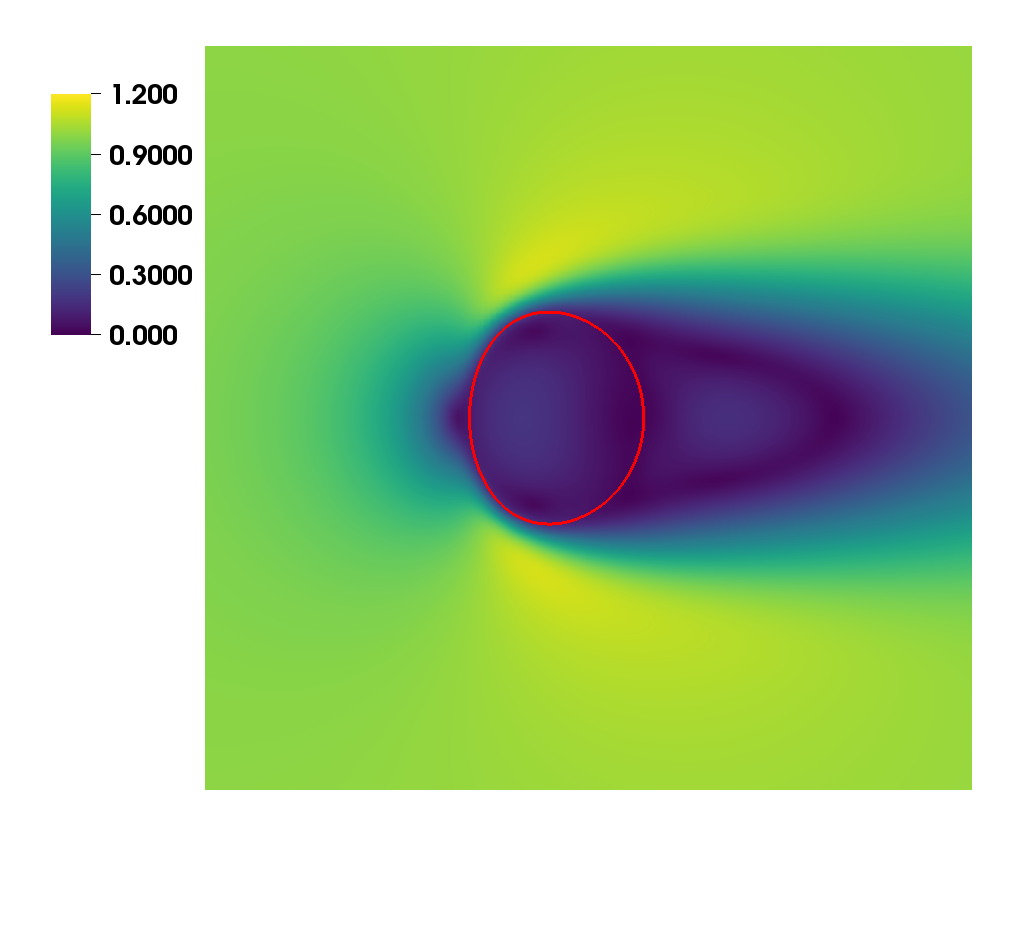}
        \caption{$We=3$}
        \label{drop-we3}
    \end{subfigure}
    \begin{subfigure}{0.4\textwidth}
    \includegraphics[width=\textwidth]{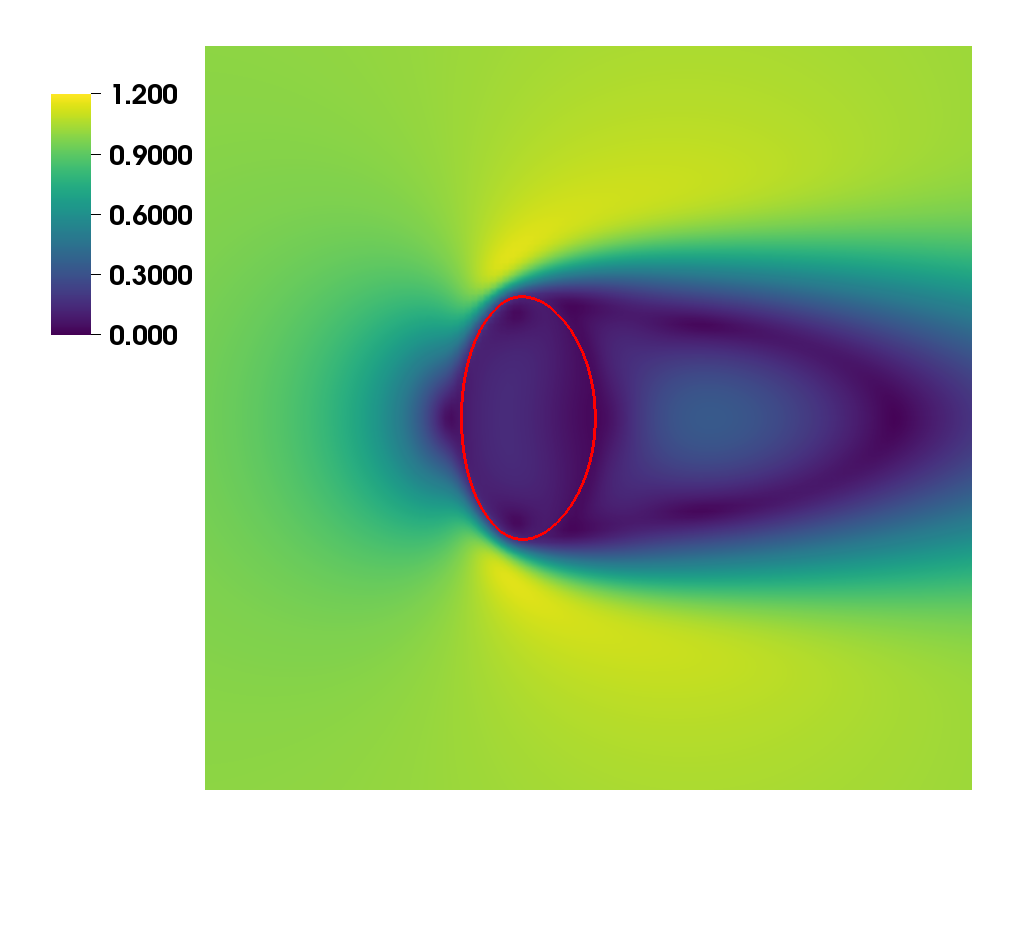}
        \caption{$We=6$}
        \label{drop-we6}
    \end{subfigure}
    \begin{subfigure}{0.4\textwidth}
    \includegraphics[width=\textwidth]{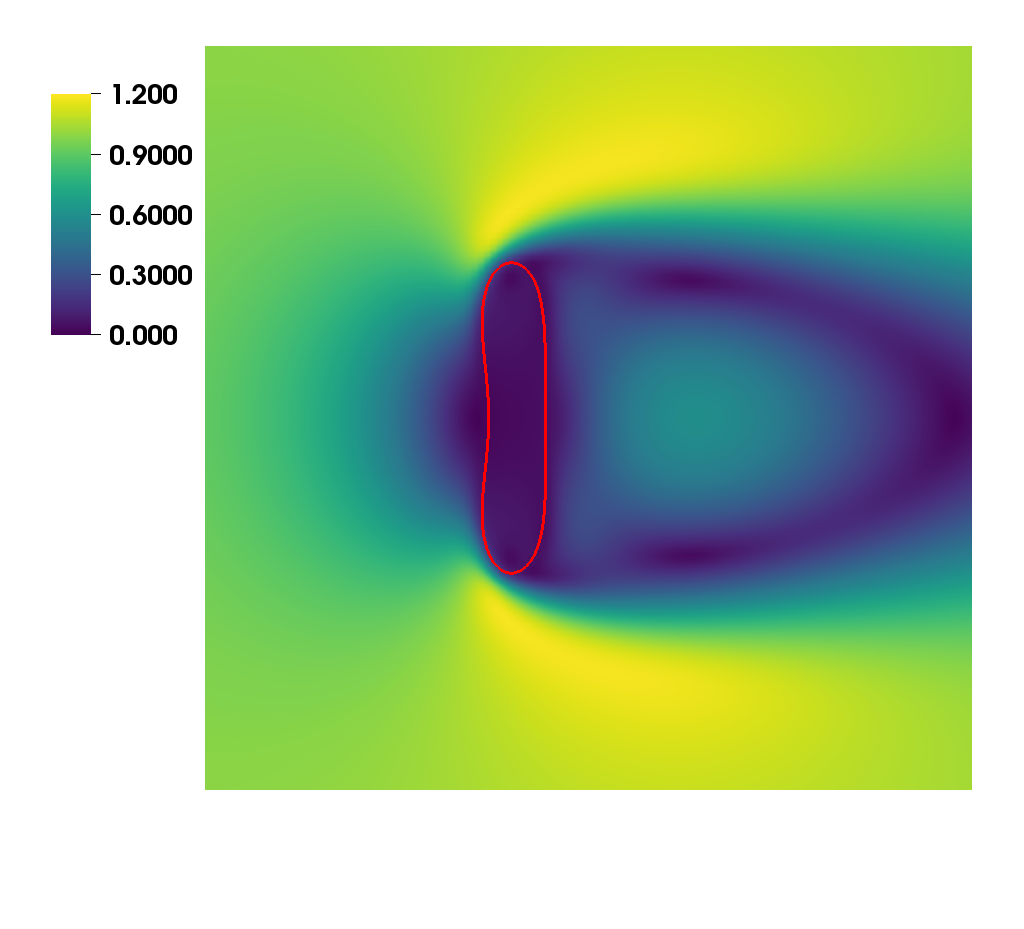}
        \caption{$We=9$}
        \label{drop-we9}
    \end{subfigure}
    \caption{Droplet deformation at different $We$ at $t^*=10$; pseudocolor is velocity magnitude normalized by inflow velocity, and red circle is liquid-gas interface}
    \label{deformation}
\end{figure}

To examine how droplet deformation will affect droplet dynamics in high pressure, we compare droplet drag coefficient with different $We$ at fixed $\rho^*=20$. We first examine whether the spheroidal deformation assumption is valid or not by comparing the droplet aspect ratio to literature. Fig.\ref{deformation} illustrates the droplet shape under different Weber number conditions. It is clear that for $We=3$ and $We=6$, droplets are still close to spheroidal shape, but for $We=9$ the droplet becomes too flat.

\subsubsection{Aspect Ratio}
In \cite{helenbrook_quasi-steady_2002}, the correlation of aspect ratio related to $We$, $Oh$, $\rho^*$ and $\mu^*$ was obtained through fully resolved simulations, while in \cite{loth_quasi-steady_2008}, the correlation was found to be dependent on $We$ only based on experiment data obtained by \citet{reinhart_verhalten_1964}. The comparison is summarized in Table.\ref{E-compare}, and it can be found that our results have good agreement with \cite{helenbrook_quasi-steady_2002,loth_quasi-steady_2008}, except for when $We=9$, which is near the onset of breakup and the deformation.

\begin{table}[ht]
\centering
\begin{tabular}{||c|c|c|c|c||} 
 \hline
 We & 1 & 3 & 6 & 9 \\  
 \hline\hline
$e$ in simulations & 0.9110 & 0.8007 & 0.5869 & 0.1743 \\ 
$e$ in \cite{helenbrook_quasi-steady_2002}& 0.8910 & 0.7336 & 0.5332 & 0.3530 \\ 
$e$ in \cite{loth_quasi-steady_2008} & 0.9178 & 0.7611 & 0.5662 & 0.4320 \\ 
$error$ to \cite{helenbrook_quasi-steady_2002} & 2.253\% & 9.157\% & 10.07\% & 50.64\% \\
$error$ to \cite{loth_quasi-steady_2008} & 0.741\% & 5.204\% & 3.652\% & 59.66\% \\
 \hline
\end{tabular}
\caption{Aspect ratio in different $We$ at $\rho^*=20$, and comparisons to \cite{loth_quasi-steady_2008} and \cite{helenbrook_quasi-steady_2002}}
\label{E-compare}
\end{table}

\begin{figure}[ht]
    \centering
    \includegraphics[width=0.6\textwidth]{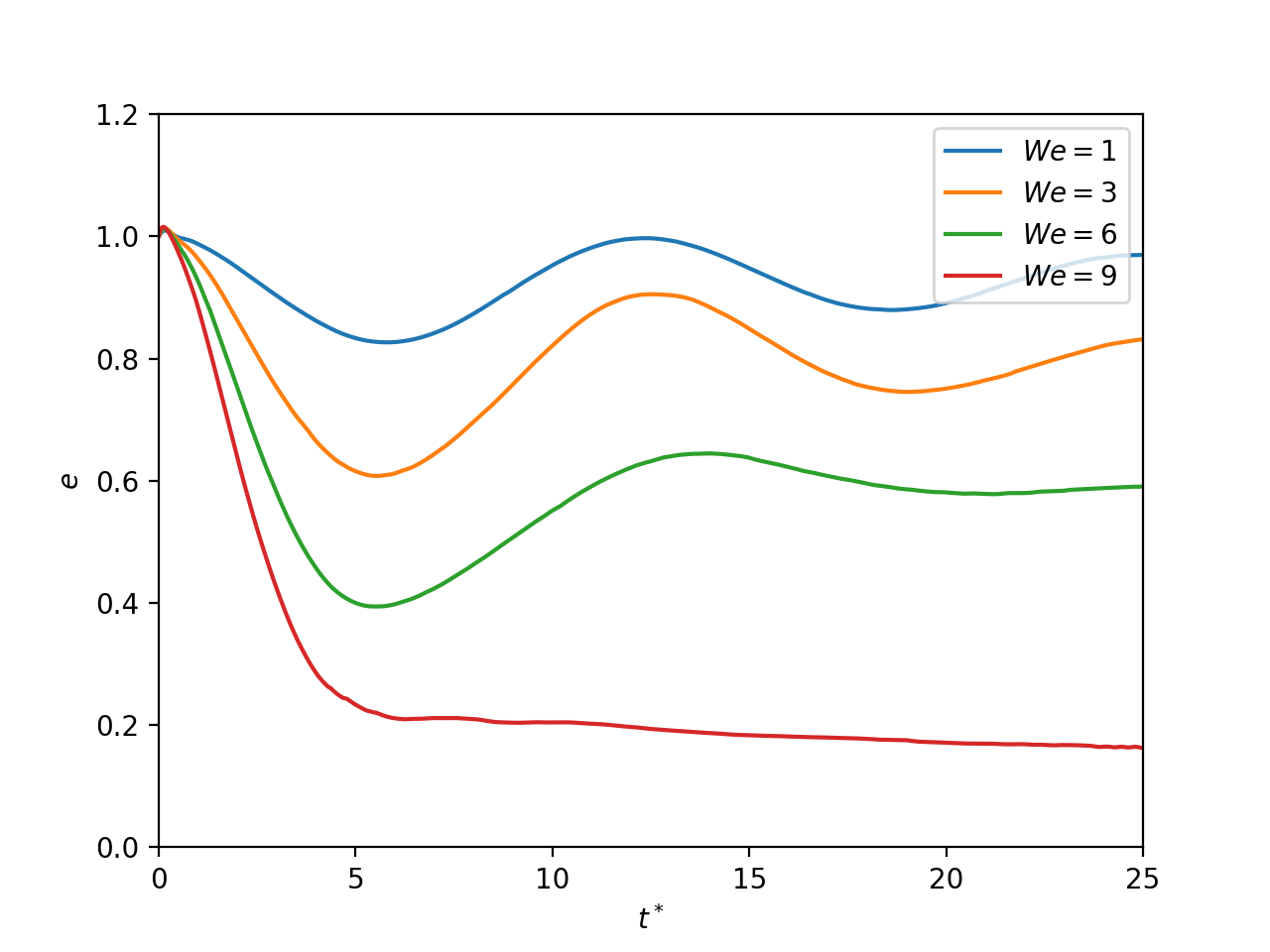}
    \caption{The evolution of droplet aspect ratio over time; $t*$ is time normalized by the capillary time scale $\tau_c$}
    \label{oscillation}
\end{figure}

The deformation has shown an oscillatory behavior, see in Fig.\ref{oscillation}. The period of oscillation are $12.5, 12.9, 13.5$ for $We=1,2,6$ respectively, which are close to theoretical prediction of $12.3$ regardless of $We$ in \cite{ashgriz_handbook_2011}. However, for $We=9$, the deformation is so strong that oscillation did not happen.
\subsubsection{Acceleration}
We compare results from different acceleration calculations using drag coefficients estimated by Eq.\ref{drag_new}. The area  $A_p$ is calculated by using $e$, as explained in the next section. When droplet reaches steady state, results with and without acceleration are very close to each other. Their lines are parallel to each other, but deviate a small value because of the slow motion of droplet. During the transient period, drag coefficients calculated with acceleration are smaller than non-accelerating cases. Acceleration calculated by $a_2=\frac{d^2X_d}{dt^2}$ oscillates very frequently when droplet enters from transient to steady state. For acceleration calculated by $a_3=\frac{dU_d^2}{2dX_d}$, the result performs poorly in transient period, because $a$ becomes too sensitive to $dX_d$. The back and forth motion of droplet due to gravity update scheme makes it not a good option for calculating acceleration. Therefore $a_1=\frac{dU_d}{dt}$ is adopted in Eq.\ref{drag_new} and will used for the rest of the paper.

\begin{figure}[ht]
    \begin{subfigure}{0.44\textwidth}
        \includegraphics[width=\textwidth]{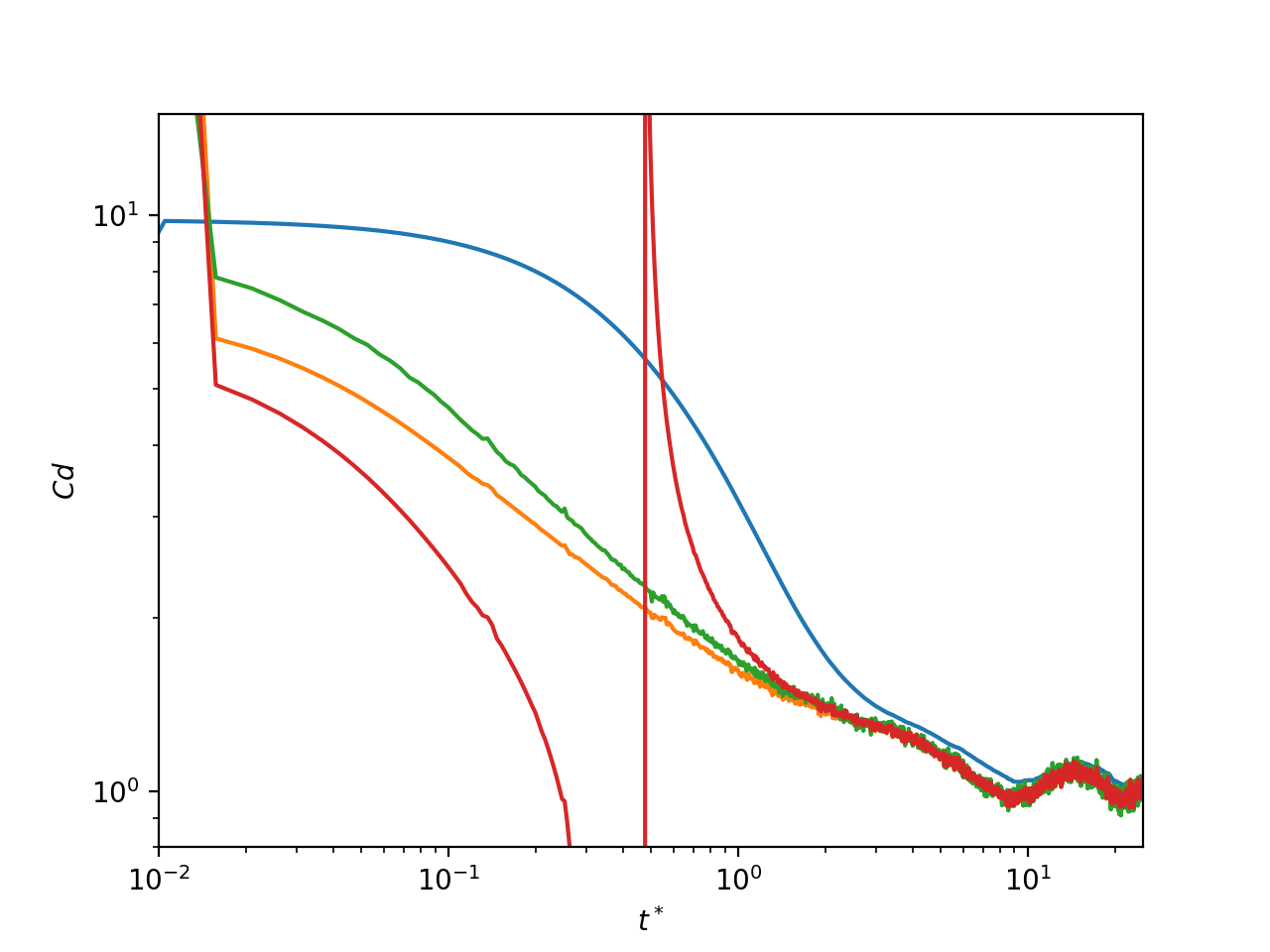}
        \caption{$We=1$}
        \label{drag-we1}
    \end{subfigure}
    \begin{subfigure}{0.44\textwidth}
    \includegraphics[width=\textwidth]{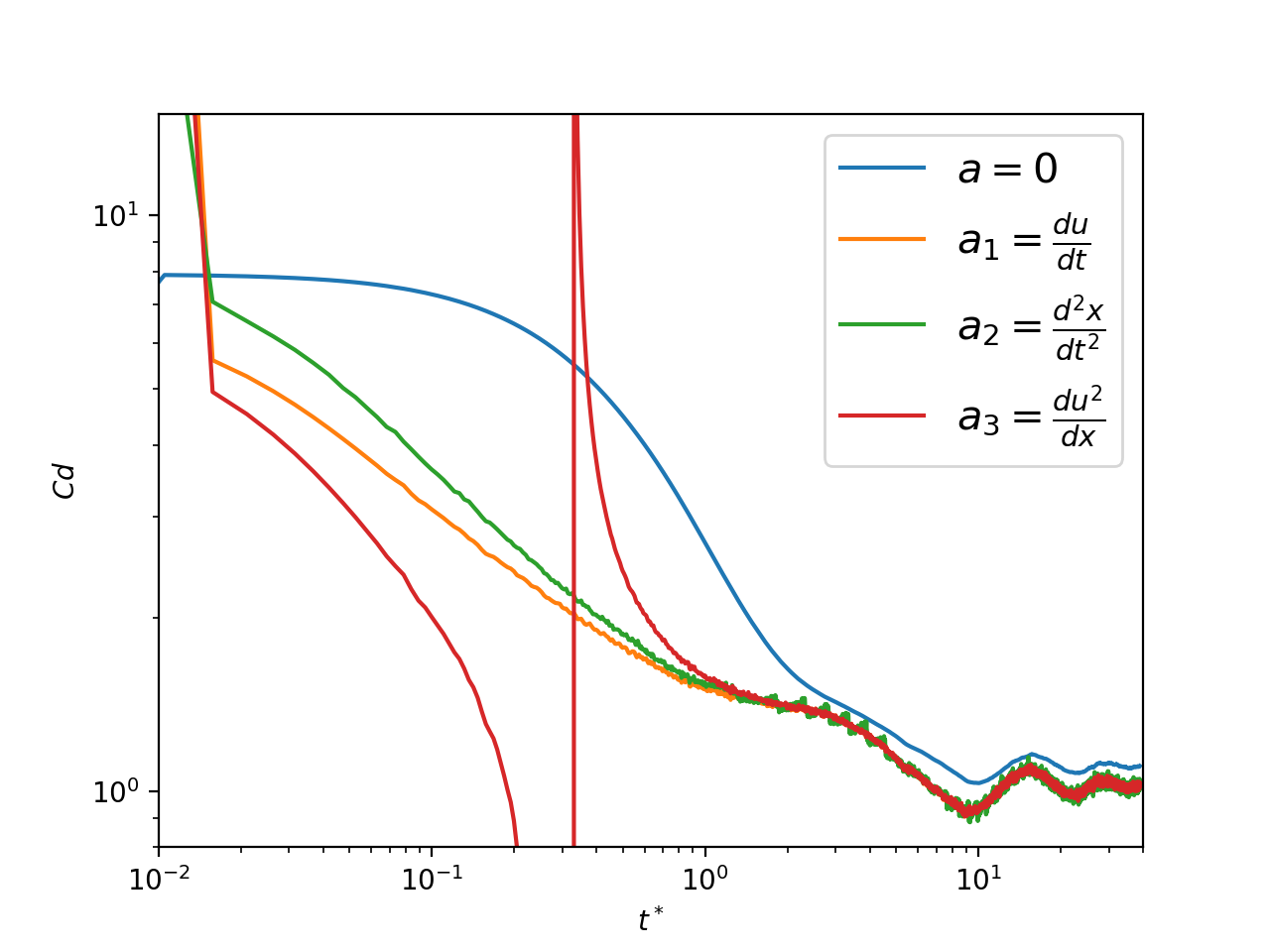}
        \caption{$We=3$}
        \label{drag-we3}
    \end{subfigure}
    \begin{subfigure}{0.44\textwidth}
    \includegraphics[width=\textwidth]{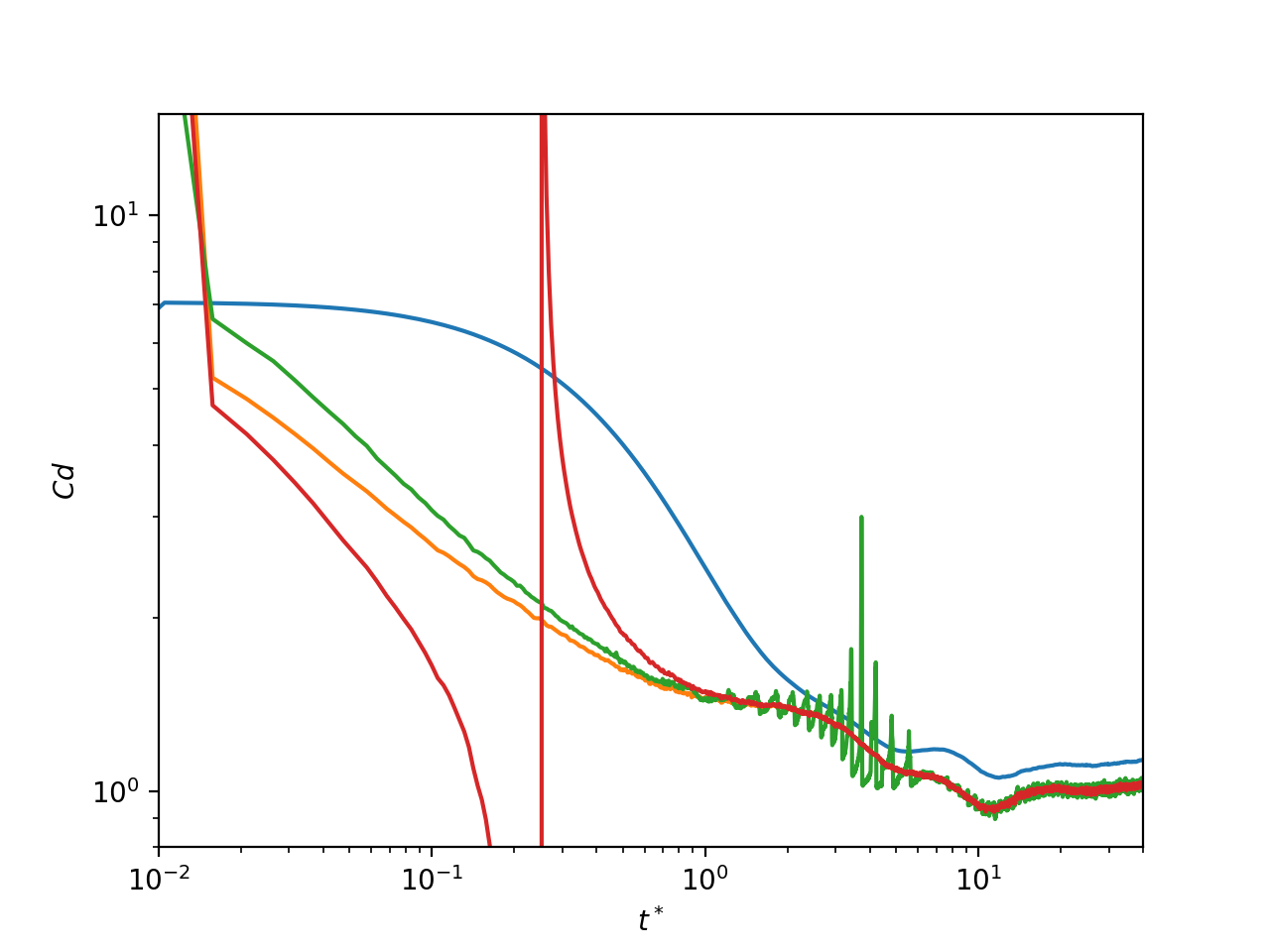}
        \caption{$We=6$}
        \label{drag-we6}
    \end{subfigure}
    \begin{subfigure}{0.44\textwidth}
    \includegraphics[width=\textwidth]{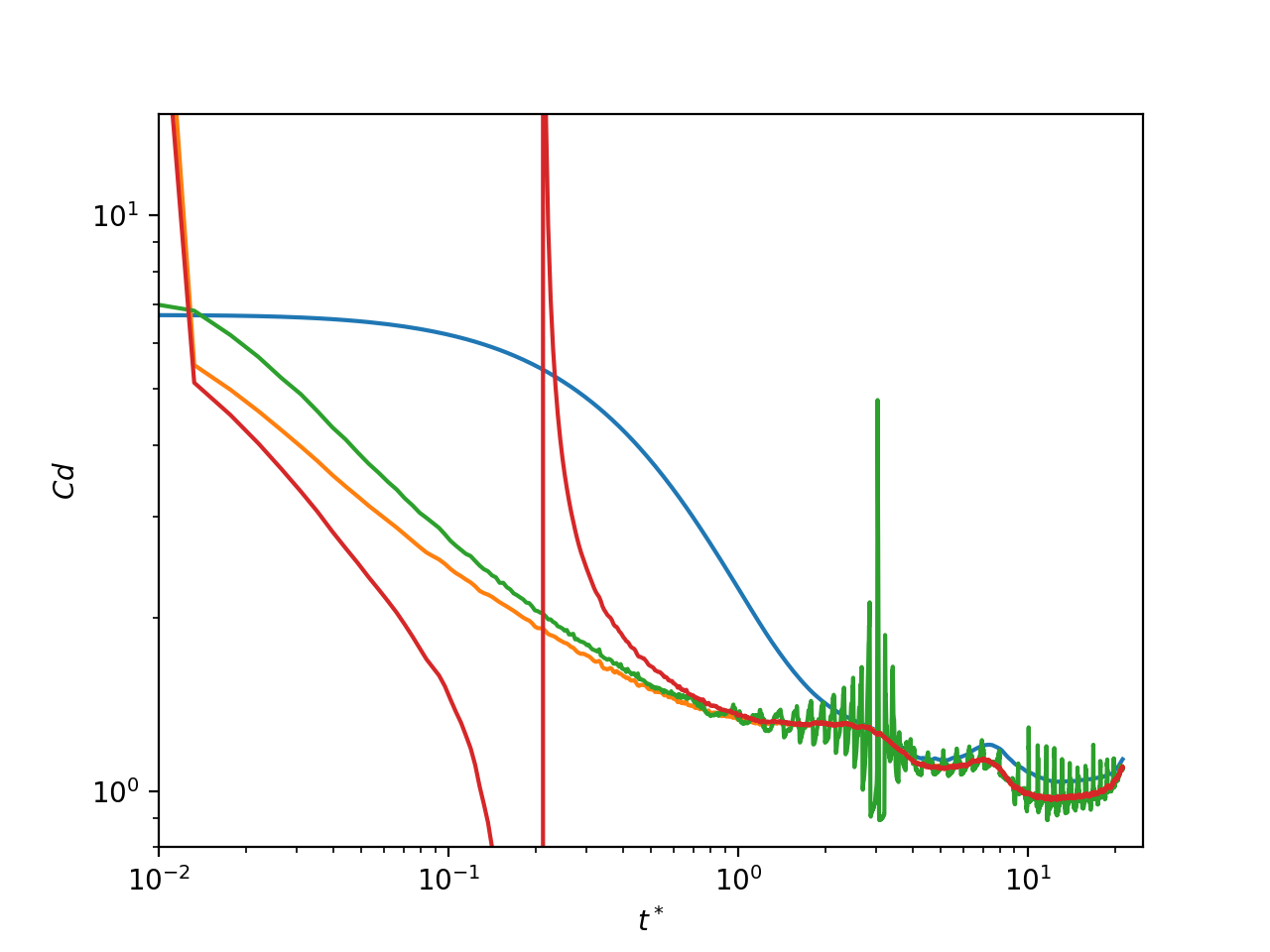}
        \caption{$We=9$}
        \label{drag-we9}
    \end{subfigure}
    \caption{Drag coefficient development for different $We$ with $a$ calculated in different ways}
    \label{drag-a}
\end{figure}
\subsubsection{Overall Drag Coefficient Prediction}
\begin{figure}[ht]
    \centering
    \includegraphics[width=0.6\textwidth]{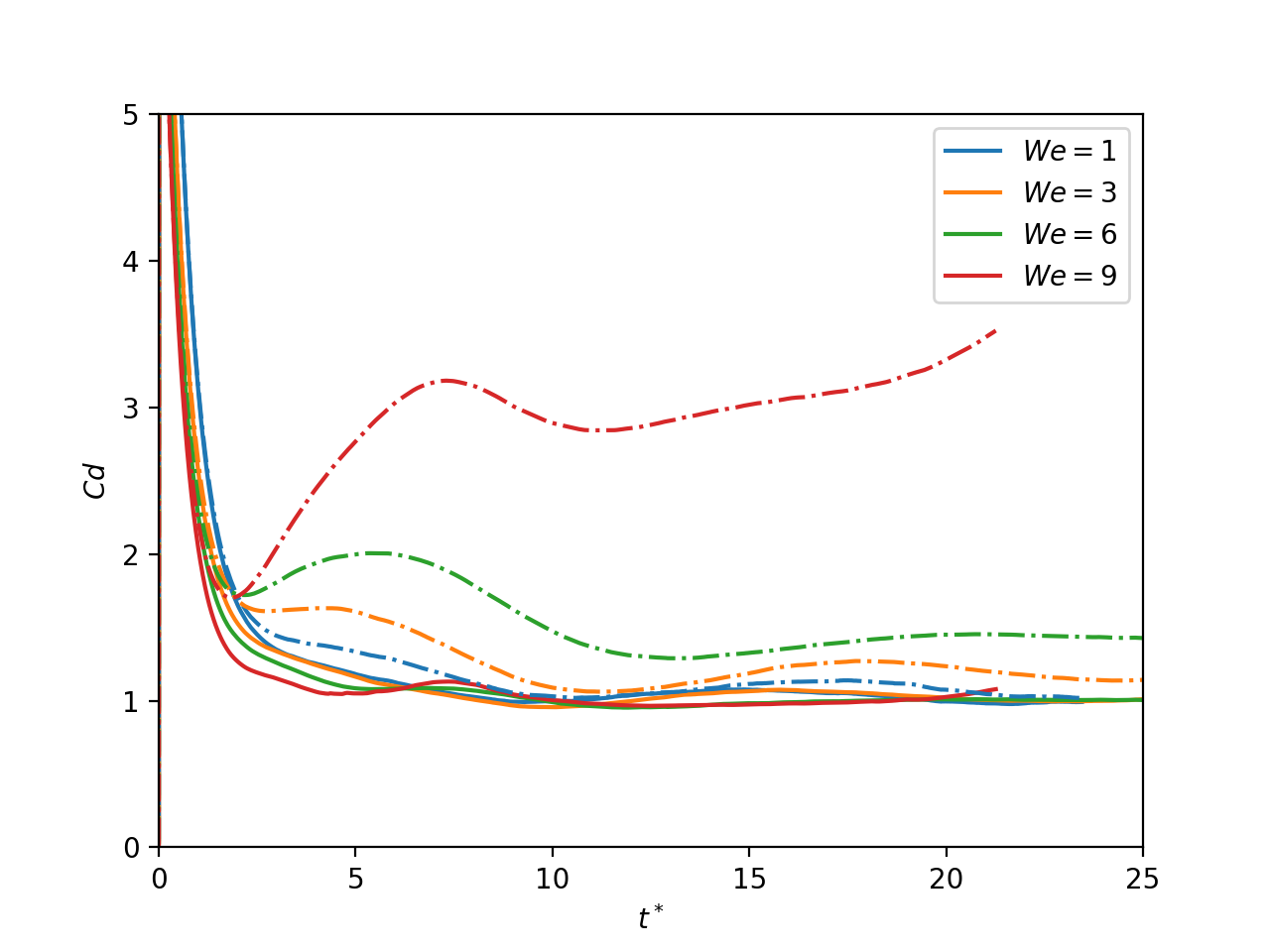}
    \caption{Drag coefficient calculated by Eq.\ref{drag_old} with $A_p$ estimated by $r_{\textrm{eff}}$ (dash-dotted lines) and $e$ (solid lines)}
    \label{drag-steady}
\end{figure}

\begin{figure}[ht]
    \centering
    \includegraphics[width=0.6\textwidth]{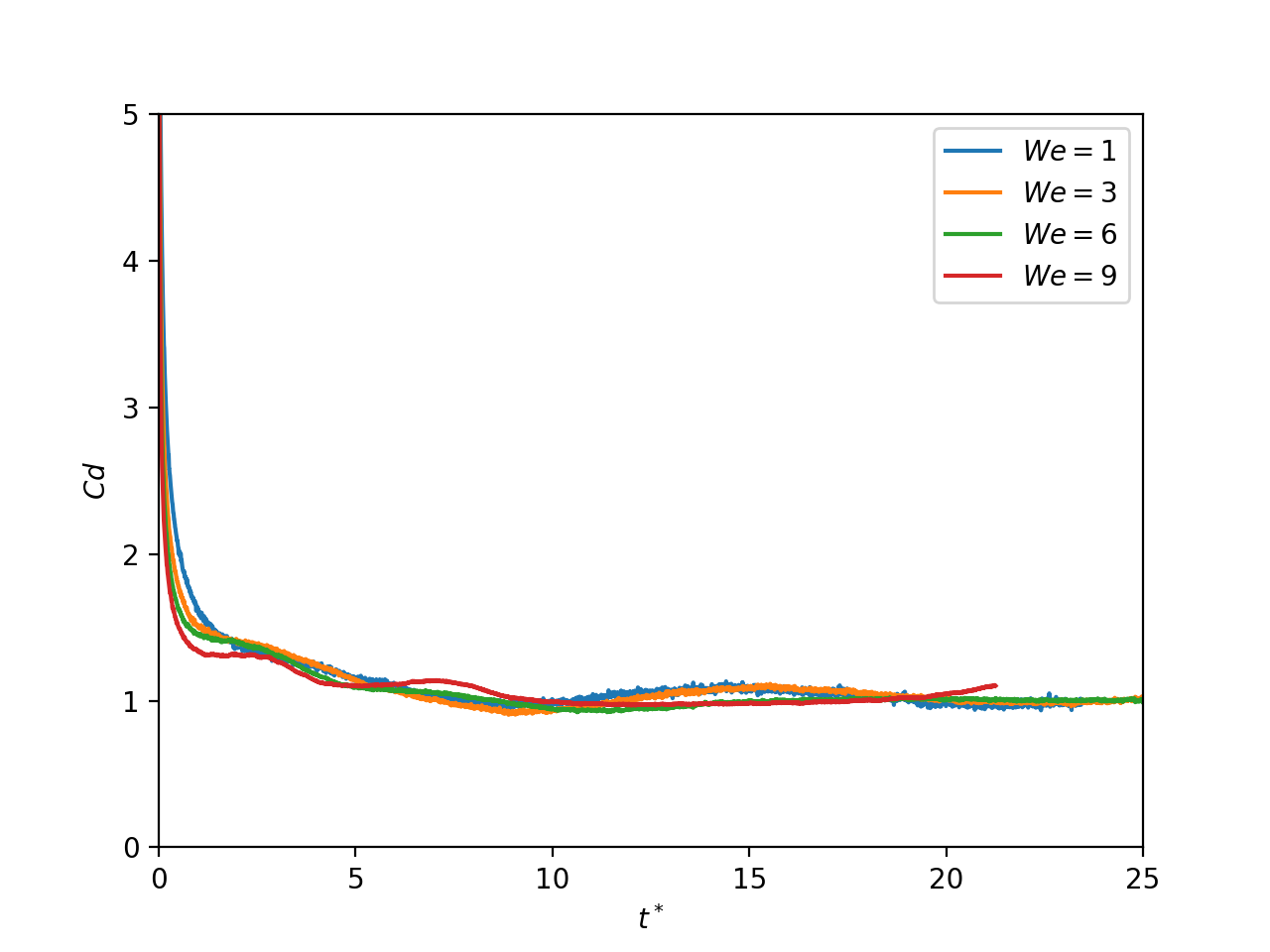}
    \caption{Drag coefficient calculated by Eq.\ref{drag_new} with $A_p$ estimated by $e$}
    \label{drag-transient}
\end{figure}

Drag coefficients calculated from Eq.\ref{drag_old} and Eq.\ref{drag_new} are plotted in Fig.\ref{drag-steady} and Fig.\ref{drag-transient} respectively. In Fig.\ref{drag-steady}, the droplets are assumed to be steady, and $C_d$ computed from different estimations of $A_p$ are presented. One of the estimation of $A_p$ is to use $r_\text{eff}$, and the other is to use the aspect ratio $e$ to update a more accurate value of the frontal area and use that instead in the area definition. $C_d$ computed by using $r_\text{rff}$ increases with increasing $We$, but this trend is mild when $C_d$ is computed by using $e$. In Fig.\ref{drag-transient}, the transient behavior of droplets are considered. The transient $C_d$ in the transient time drops faster than the $C_d$ computed by Eq.\ref{drag_old}.


\begin{figure}[ht]
    \centering
    \includegraphics[width=0.6\textwidth]{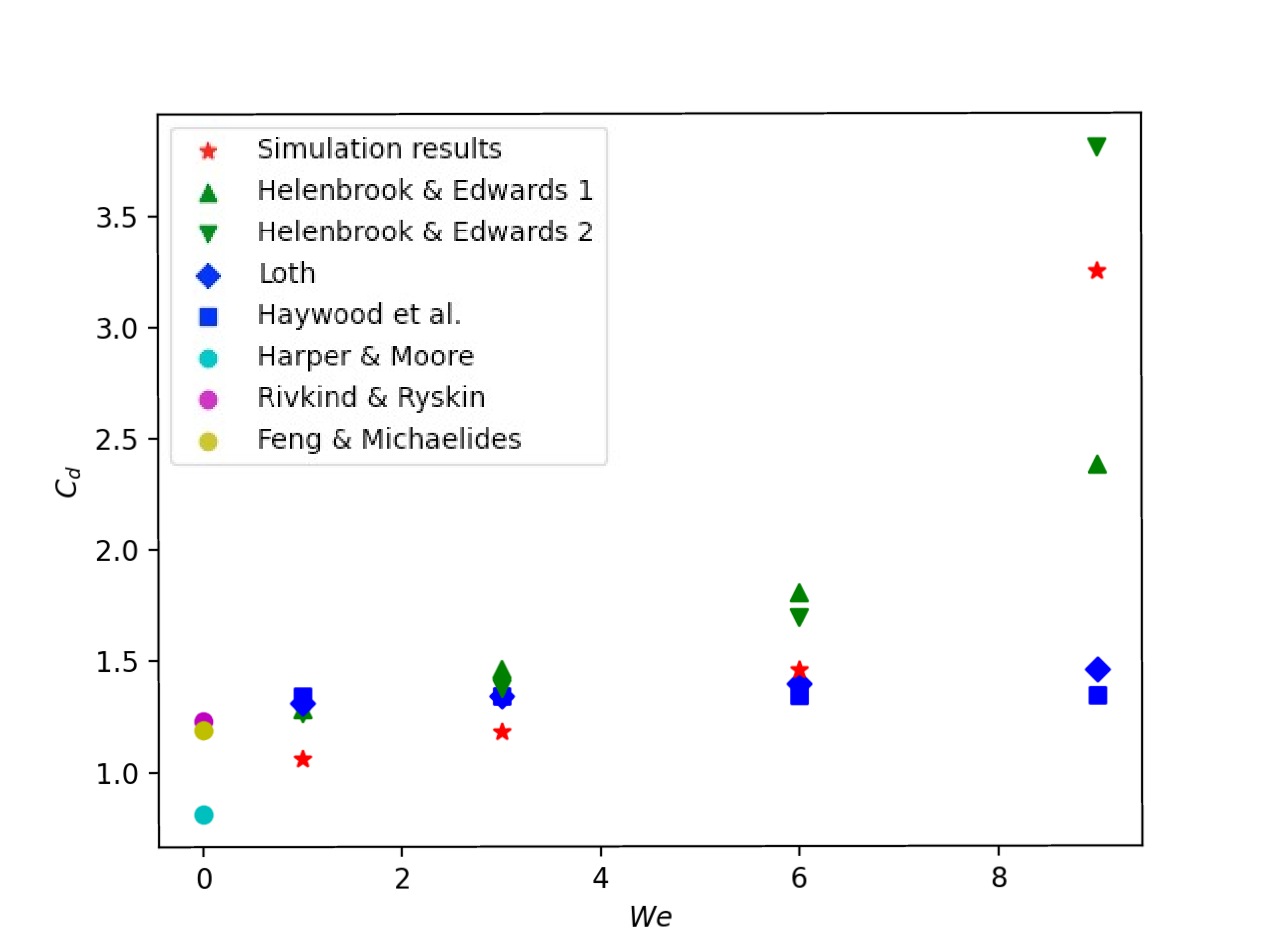}
    \caption{Comparison of drag coefficient to literature}
    \label{compareWe}
\end{figure}

In addition, comparison of drag coefficient to literature have been made, shown in Fig.\ref{compareWe}. There are two sets of data for \citet{helenbrook_quasi-steady_2002}. For green upward triangle data points, $C_d$ and $e$ are both obtained using correlations in \cite{helenbrook_quasi-steady_2002}. For green downward triangle points, $e$ is replaced by our simulation results. Blue data points are from correlations in \cite{haywood_numerical_1994,loth_quasi-steady_2008} for deformable liquid droplets, while round data points are only for spherical liquid droplets from \cite{harper_motion_1968,rivkind_flow_1976,feng_drag_2001}. Helenbrook and Edwards' work considers both deformation and internal circulation. All drag coefficient correlations from literature use $r_{\text{eff}}$ to calculate $C_d$, thus we use drag coefficient calculated by Eq.\ref{drag_old} for comparison.

In Fig.\ref{compareWe}, we see our data in low $We$ are quite close to others' work. But at $We=9$, since the droplet is highly deformed, the shape of droplet cannot be treated as spheroid. As a result, the evaluation of $e$ is not accurate. Besides, results from literature have no agreement at $We=9$, meaning predictions of $C_d$ in current literature perform bad when the droplet is highly deformed and near breakup.

\subsection{Changing $\rho^*$ at fixed $We$}
\begin{figure}[ht]
    \begin{subfigure}{0.48\textwidth}
        \includegraphics[width=\textwidth]{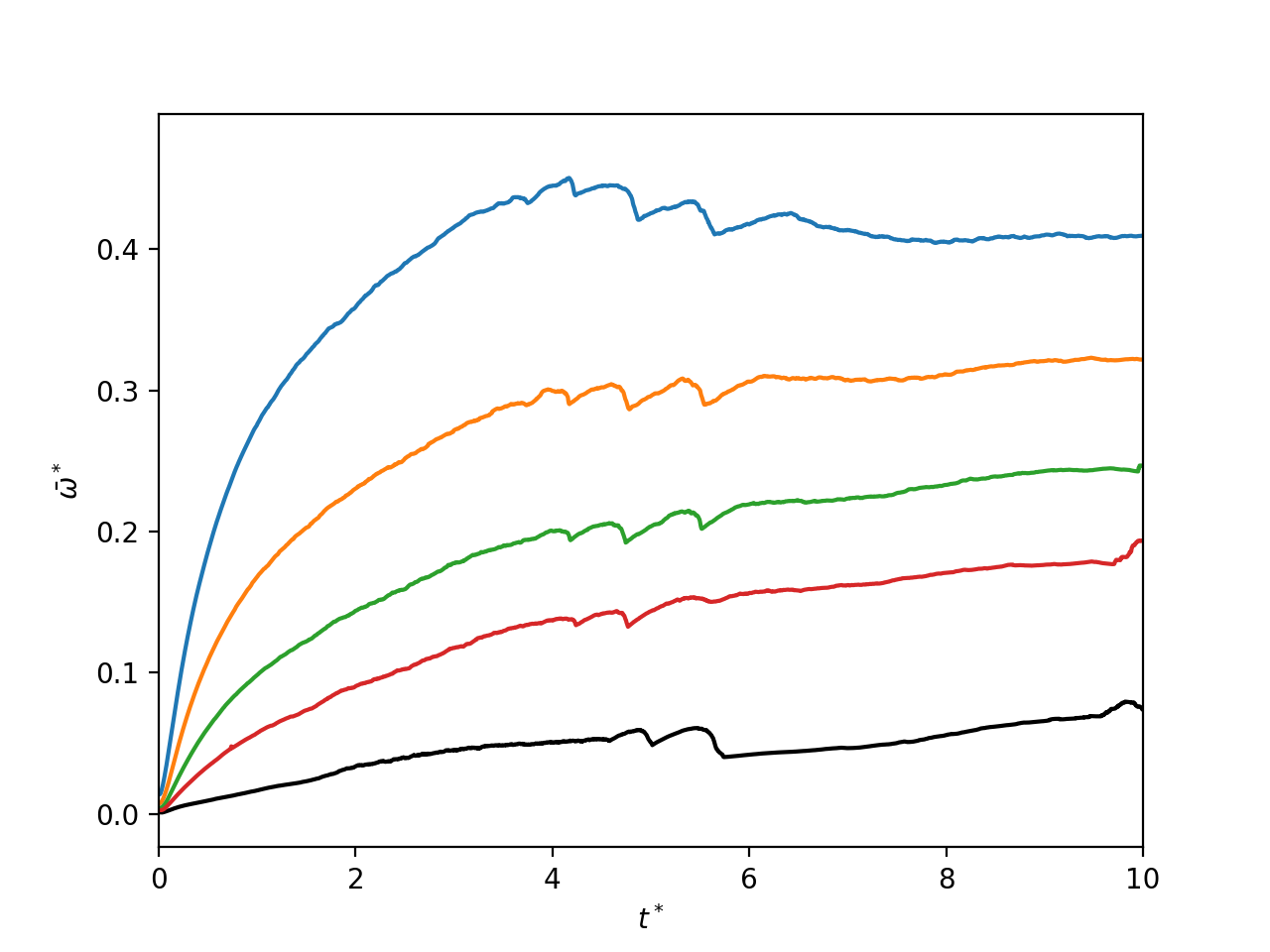}
        \caption{$\bar{\omega}^*$ vs $t^*$}
        \label{ommeanwe6}
    \end{subfigure}
    \begin{subfigure}{0.48\textwidth}
    \includegraphics[width=\textwidth]{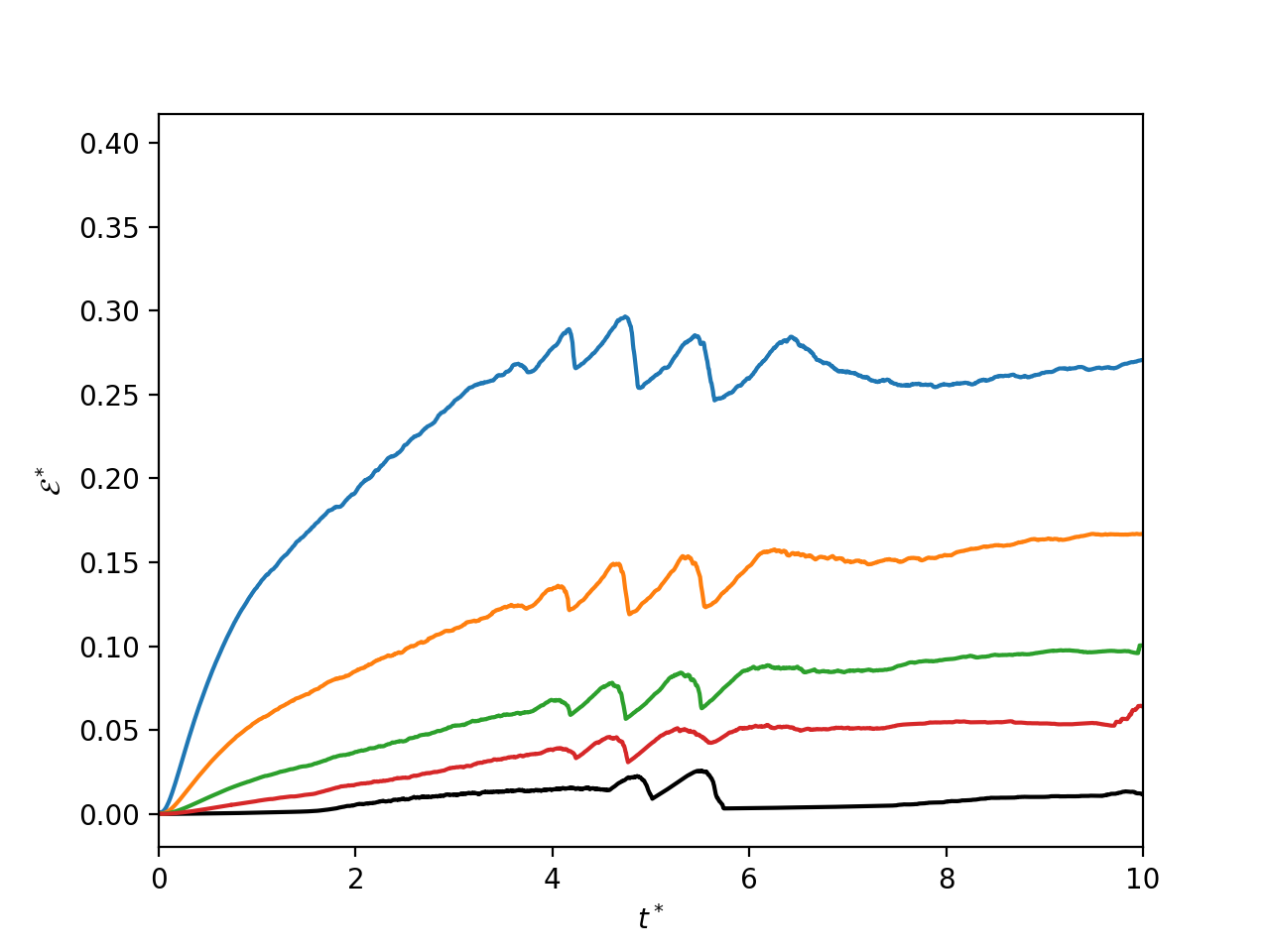}
        \caption{$\bar{\mathcal{E}}^*$ vs $t^*$}
        \label{enstrophywe6}
    \end{subfigure}
    \begin{subfigure}{0.48\textwidth}
    \includegraphics[width=\textwidth]{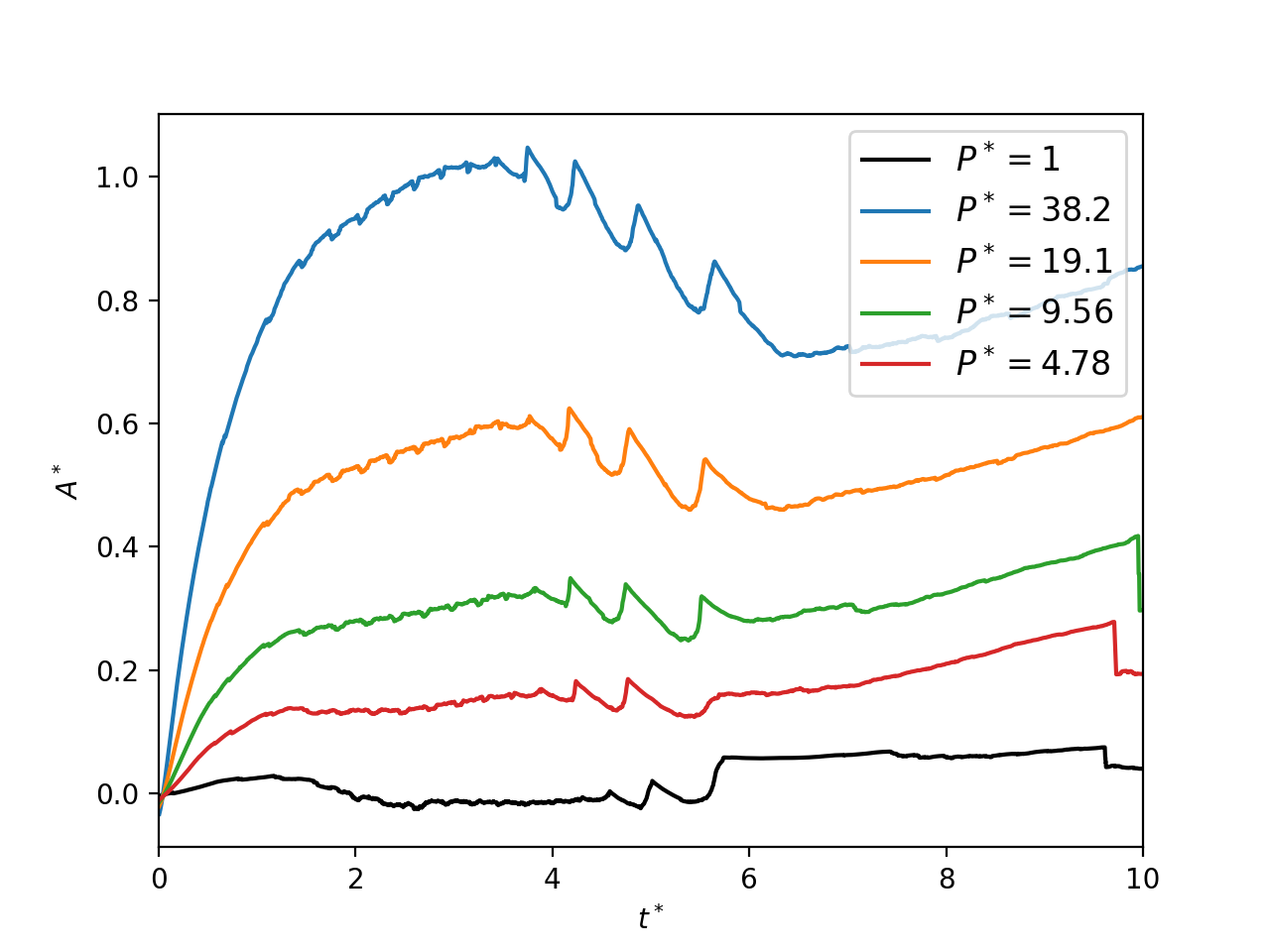}
        \caption{$\bar{A}^*$ vs $t^*$}
        \label{hillwe6}
    \end{subfigure}
    \caption{Correlation between droplet internal circulation and pressure; the volume-averaged vorticity, enstrophy and Hill's constant are nondimensionalized by: $\bar{\omega}^*=\bar{\omega}/\left(U_{in}/D\right)$, $\bar{\mathcal{E}}^*=\bar{\mathcal{E}}/\left(0.5\left(U_{in}/D\right)^2\right)$, $\bar{A}^*=\bar{A}/\left(U_{in}/D^2\right)$}
    \label{circulation-t}
\end{figure}

$\rho^*$ is the quantification of internal circulation strength \cite{lin_numerical_2022}, and is correlated to pressure through ideal law, indicating the coupling between liquid and gas phases. This correlation can be found in Fig.\ref{circulation-t}. With increasing $P^*$, i.e. decreasing $\rho^*$, the internal circulating will become stronger regardless of the method used for measuring internal circulation strength. 

\begin{figure}[ht]
    \begin{subfigure}{0.32\textwidth}
    \includegraphics[width=\textwidth]{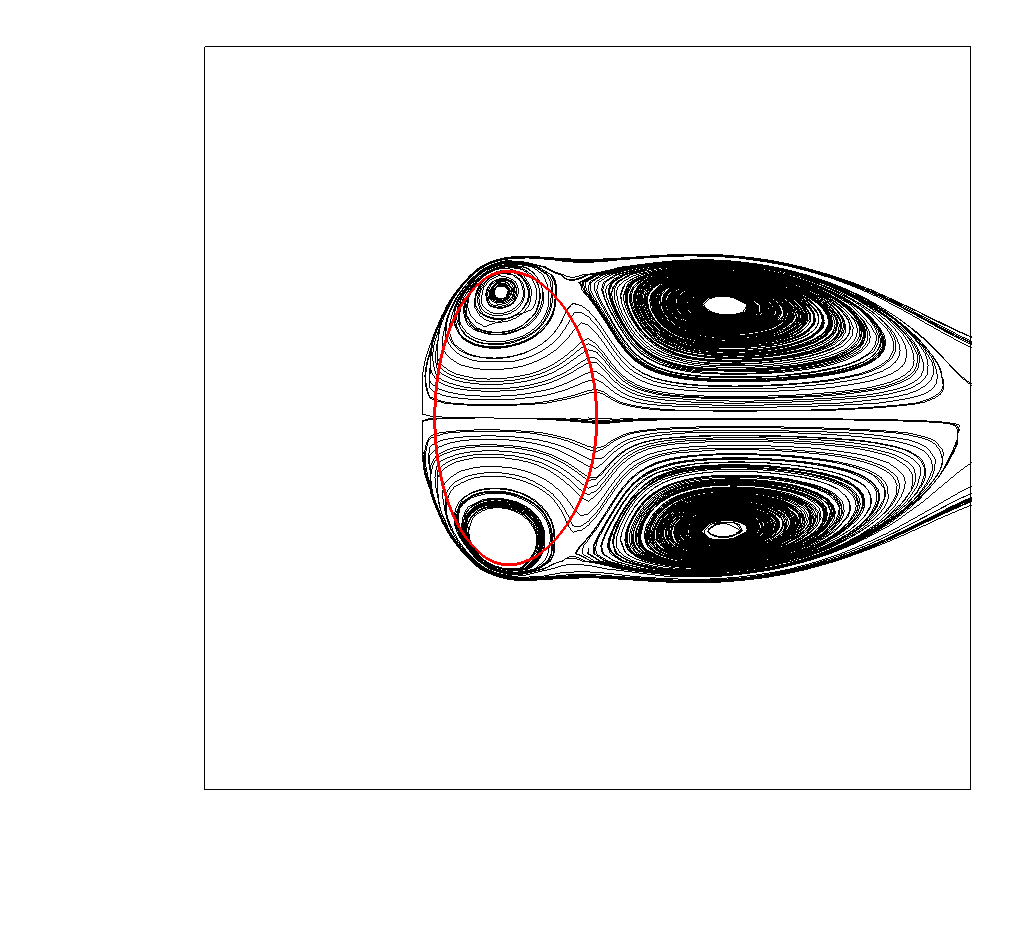}
        \caption{$\rho^*=20$}
        \label{drop-rho20}
    \end{subfigure}
    \begin{subfigure}{0.32\textwidth}
    \includegraphics[width=\textwidth]{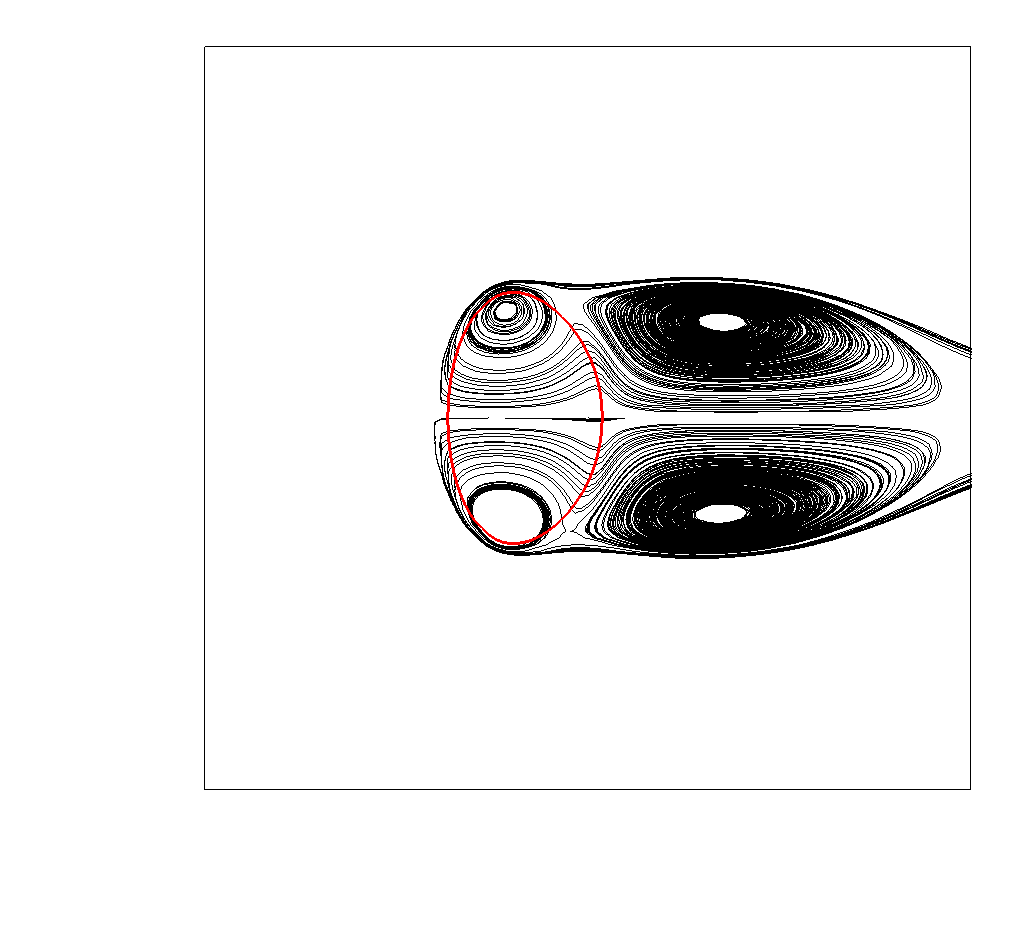}
        \caption{$\rho^*=40$}
        \label{drop-rho40}
    \end{subfigure}
    \begin{subfigure}{0.32\textwidth}
    \includegraphics[width=\textwidth]{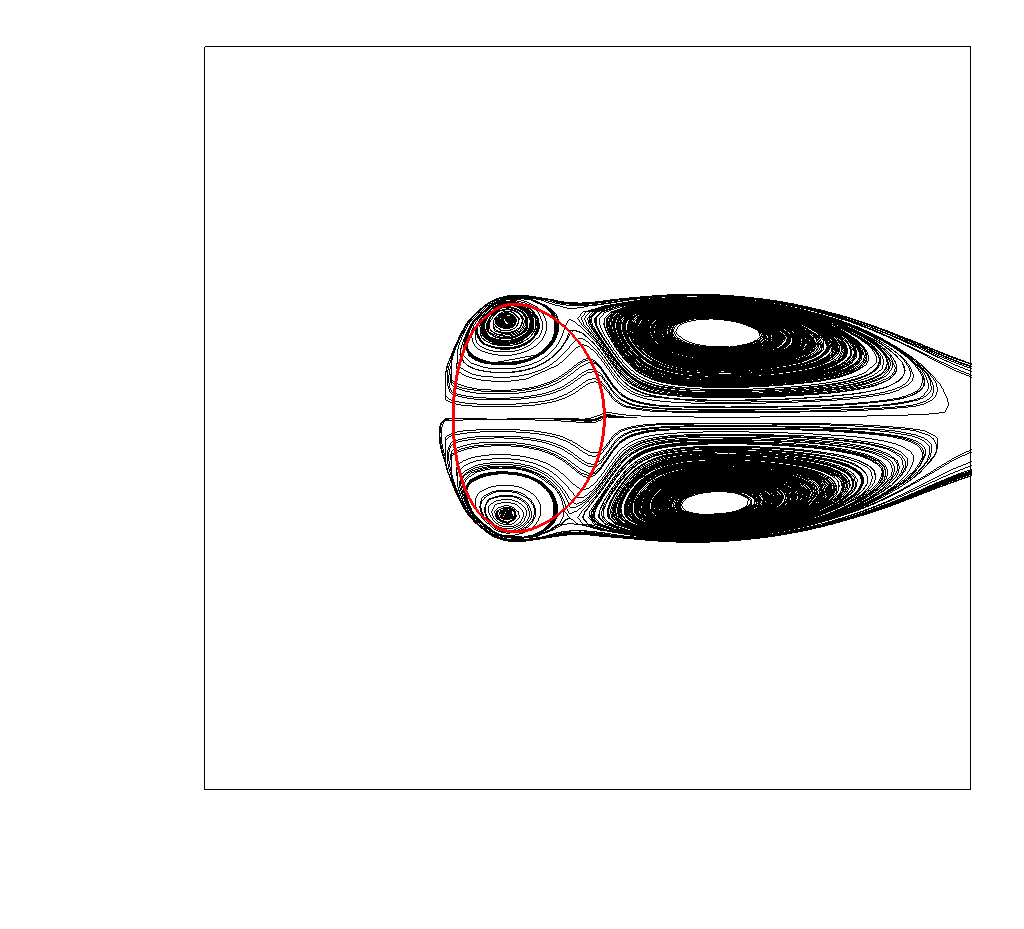}
        \caption{$\rho^*=80$}
        \label{drop-rho80}
    \end{subfigure}
    \begin{subfigure}{0.32\textwidth}
    \includegraphics[width=\textwidth]{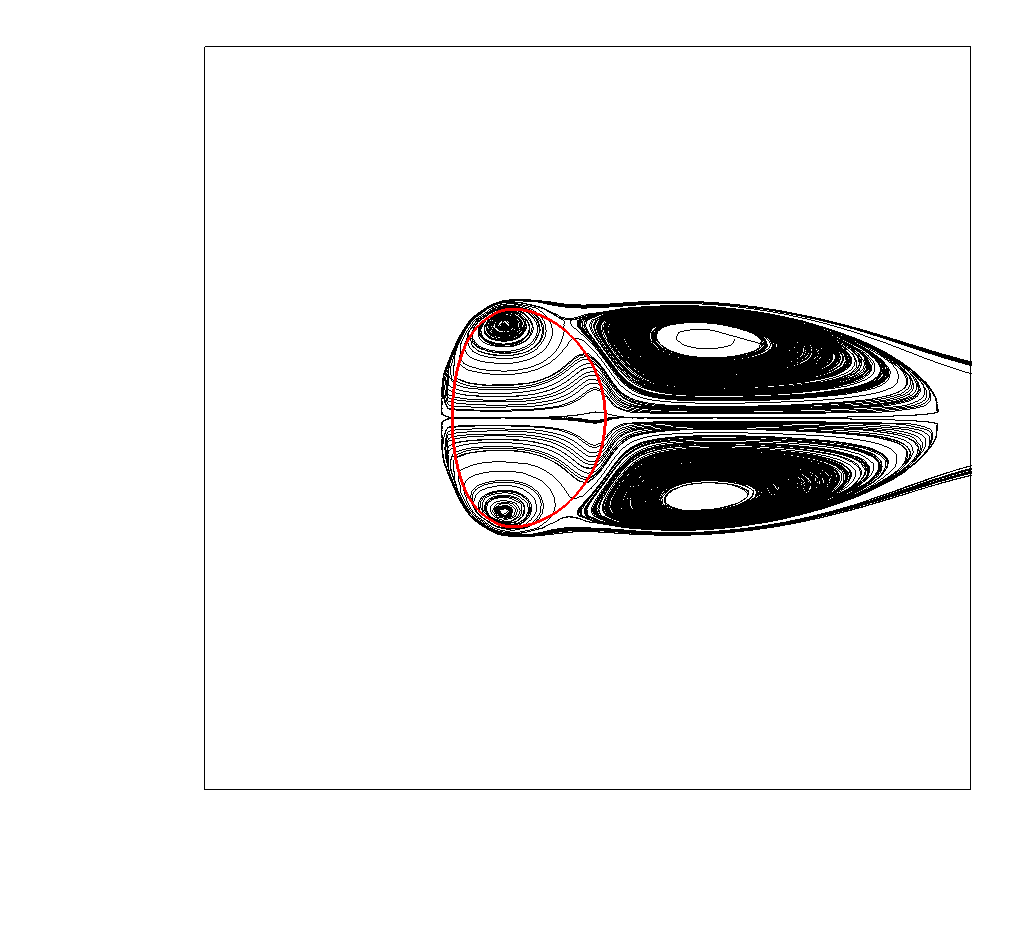}
        \caption{$\rho^*=160$}
        \label{drop-rho160}
    \end{subfigure}
    \begin{subfigure}{0.32\textwidth}
    \includegraphics[width=\textwidth]{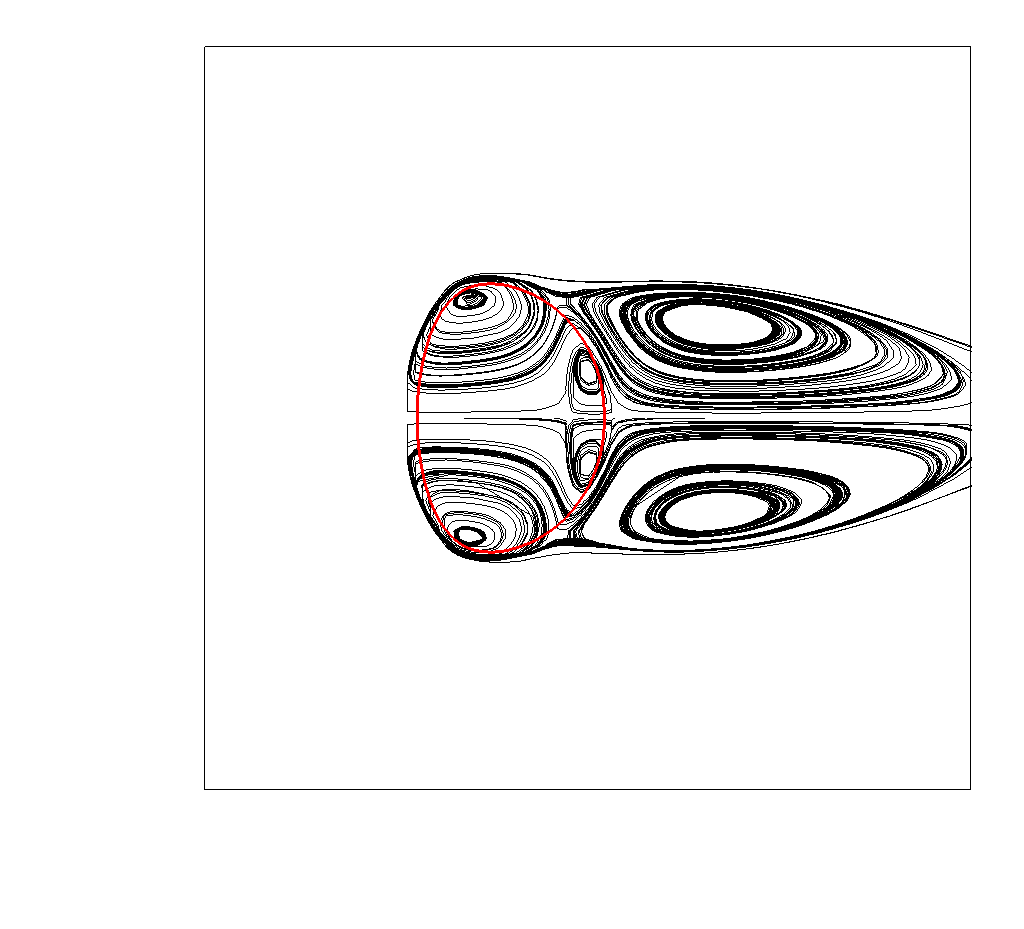}
        \caption{$\rho^*=765$}
        \label{drop-we9}
    \end{subfigure}
    \caption{Streamlines around and inside droplet at $We=6$ and $t^*=10$ with different $\rho^*$}
    \label{internalflow}
\end{figure}

\begin{figure}
    \centering
    \includegraphics[width=0.6\textwidth]{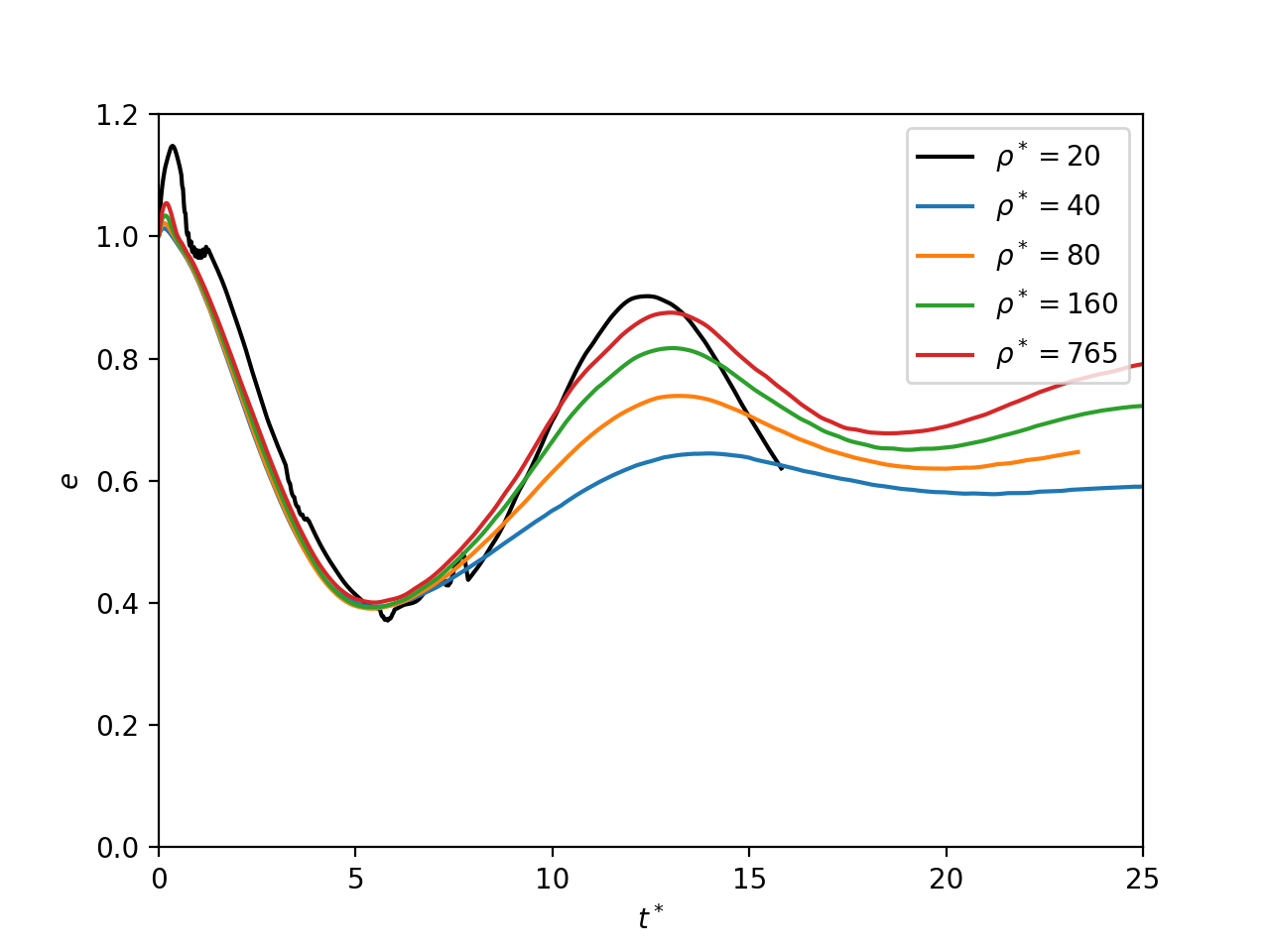}
    \caption{The evolution of droplet aspect ratio over time}
    \label{oscillation-we}
\end{figure}

To consider solely the effect of internal circulation, we compare drag coefficient in different $\rho^*$ at fixed $We=6$. Streamlines of droplet internal flow at $We=6$ have been shown in Fig.\ref{internalflow}. Strong circulation can be observed within the droplet. For droplet with very high liquid-gas density ratio, a secondary vortex can be found at the rear of droplet. Since the secondary vortex circulates in the orientation opposite to the primary vortex, it reduces the overall internal circulation strength \cite{ayyaswamy_effect_1990}. The evolution of droplet shape is shown in Fig.\ref{oscillation-we}. The onset of oscillation of deformation has been observed to have slight phase shift due to internal circulation difference in each case \cite{mashayek_nonlinear_1998}.

\begin{figure}
    \centering
    \includegraphics[width=0.6\textwidth]{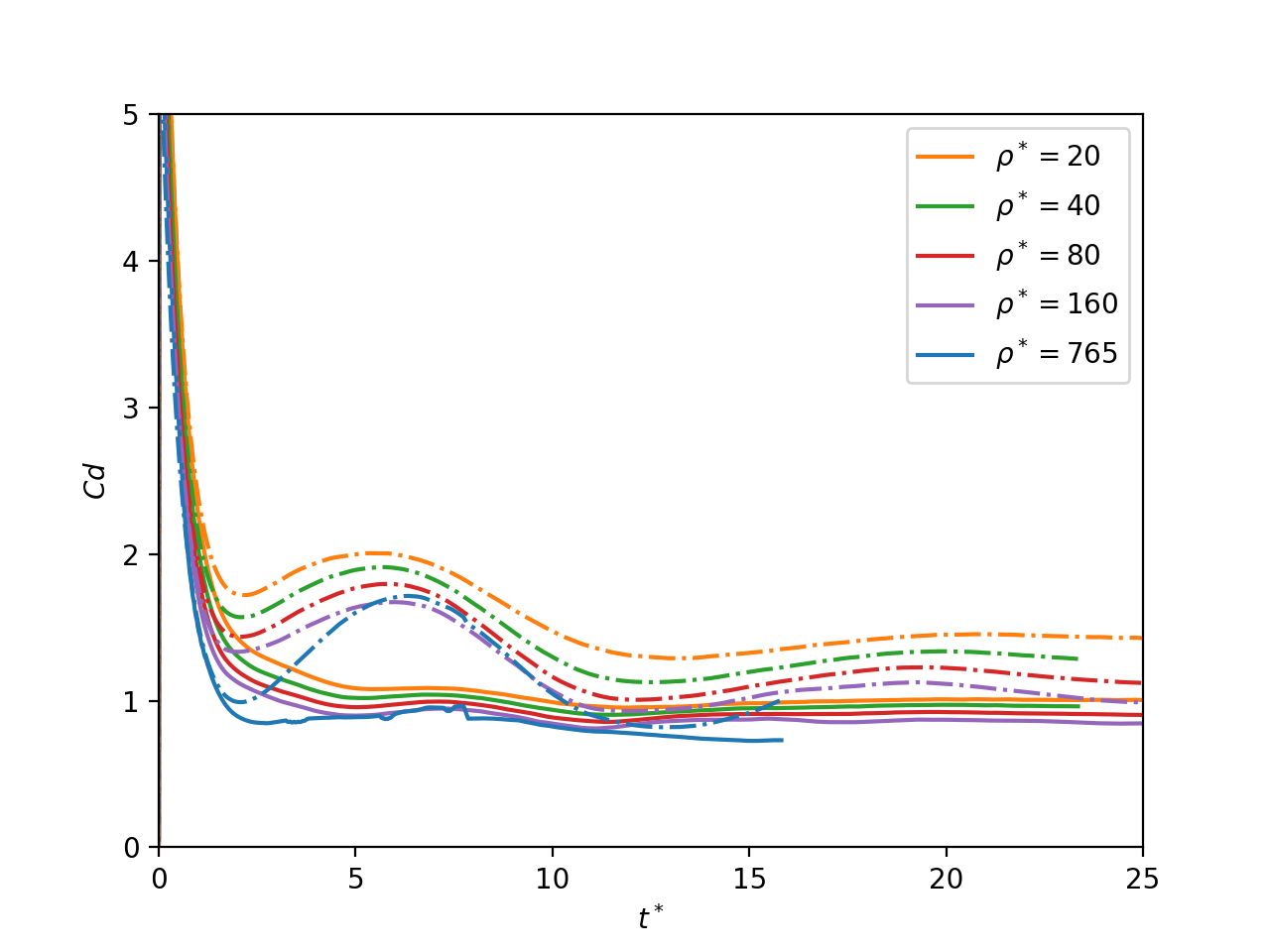}
    \caption{Drag coefficient calculated by Eq.\ref{drag_old} with $A_p$ estimated by $r_{\textrm{eff}}$ (dash-dotted lines) and $e$ (solid lines)}
    \label{drag-pressure}
\end{figure}
 
 The steady-state drag coefficients with different $\rho^*$ are plotted in Fig.\ref{drag-pressure}. Same in Sec.\ref{result-We}, $C_d$ estimated by different $A_p$ are presented. A clear trend is shown in the figure that decreasing density ratio corresponds to larger drag coefficient. This indicates that the enhanced liquid and gas coupling due to high pressure will increase drag coefficient, although it is noted that with corrected computation of $A_p$, the variation of $C_d$ becomes more mild.
 
\subsection{Scaling analysis on droplet aspect ratio}

We are concerned with what parameters affect the deformation of a droplet in a uniform convective flow for small Weber number cases. Some assumptions are made to simplify the problem. Firstly, the Ohnesorge number ($Oh$) is smaller than 0.1 in the case we are considering. Thus the viscous effect can be neglected \cite{guildenbecher_secondary_2009}. Secondly, the deformed droplet is a spheroid. More specifically, it is an oblate spheroid.

For a spherical droplet, due to the shear force on the droplet interface, there will be internal circulation inside the droplet. Therefore, we assume the gas inertial energy $E_{i,g}$ will be converted into the inertial energy of liquid $E_{i,l}$, and also provide the energy change in surface energy $E_\gamma$:

\begin{equation}\label{eqn}
    E_{i,g}\sim E_{i,l}+E_\gamma.
\end{equation}

The initial spherical droplet has a diameter $D$, and after the deformation, it becomes a spheroid with semi-major axis length and semi-minor axis length to be $R_y$ and $R_x$ respectively. The aspect ratio is then calculated as $e=R_x/R_y$, which will be smaller than $1$ in our case. A schematic of the spherical and deformed droplets are drawn in Fig.\ref{coordinate}.

The inertial energy of gas and liquid are:
\begin{equation}\label{Eig}
    E_{i,g}\sim\frac{1}{2}\rho_gU_{in}^2V,
\end{equation}
\begin{equation}\label{Eil}
    E_{i,l}\sim\frac{1}{2}\rho_lU_s^2V,
\end{equation}
where $U_s$ is the liquid velocity at droplet surface. From the definition of $e$ and conservation of mass, the relation between $D$ and $R_x$ can be found:
\begin{equation}\label{exy}
    e=\frac{R_x}{R_y}\Rightarrow R_y^2=\frac{R_x^2}{e^2},
\end{equation}
\begin{equation}\label{xD}
    V=\frac{4}{3}\pi R_xR_y^2=\frac{1}{6}\pi D^2\Rightarrow R_x=\frac{1}{2}D\cdot e^{2/3}.
\end{equation}

The surface energy change is the difference of surface energy between spherical and spheroidal shape:
\begin{equation}\label{Egamma}
    E_\gamma=\gamma\left(A_{\text{spheroid}}-A_{\text{sphere}}\right),
\end{equation}
where $A_{\text{spheroid}}=2\pi R_y^2+\pi\frac{R_x^2}{\alpha}\ln\left(\frac{1+\alpha}{1-\alpha}\right)$ with $\alpha=1-e^2$.

Usually, the freestream velocity $U_{in}$ is known, but the droplet surface velocity $U_s$ is unknown. To establish a correlation between $U_{in}$ and $U_s$, the scaling analysis based on the continuity of shear stress at the phase interface in \cite{law_theory_1977} for spherical droplets gives:
\begin{equation}\label{UsUinf}
    \frac{U_s}{U_{in}}=\left(\rho^*\mu^*\right)^{-1/3}.
\end{equation}

Using the above equations and relations, the correlation between Weber number and aspect ratio can be found:
\begin{equation}\label{relation}
    We\cdot\left[1-\left(\rho^*/\mu^{*2}\right)^{1/3}\right]\sim2\left(e^{-2/3}-2\right)+\frac{e^{4/3}}{\sqrt{1-e^2}}\ln\left(\left(\frac{1+\sqrt{1-e^2}}{e}\right)^2\right).
\end{equation}

If the deformation is very small, we can further simplify Eq.\ref{relation} by using Taylor expansion on the RHS at $e=1$, and eliminating higher order terms to get:
\begin{equation}\label{simplify}
    \left(1-e\right)^2\sim We\cdot\left(1-\left(\rho^*/\mu^{*2}\right)^{1/3}\right).
\end{equation}
Eq.\ref{simplify} has a very limited application, because from the observation in our numerical works, the spheroidal assumption can only hold at the range around $0.5<e<1$.

\begin{figure}[!ht]
    \begin{subfigure}{0.49\textwidth}
    \centering
    \includegraphics[width=1\textwidth]{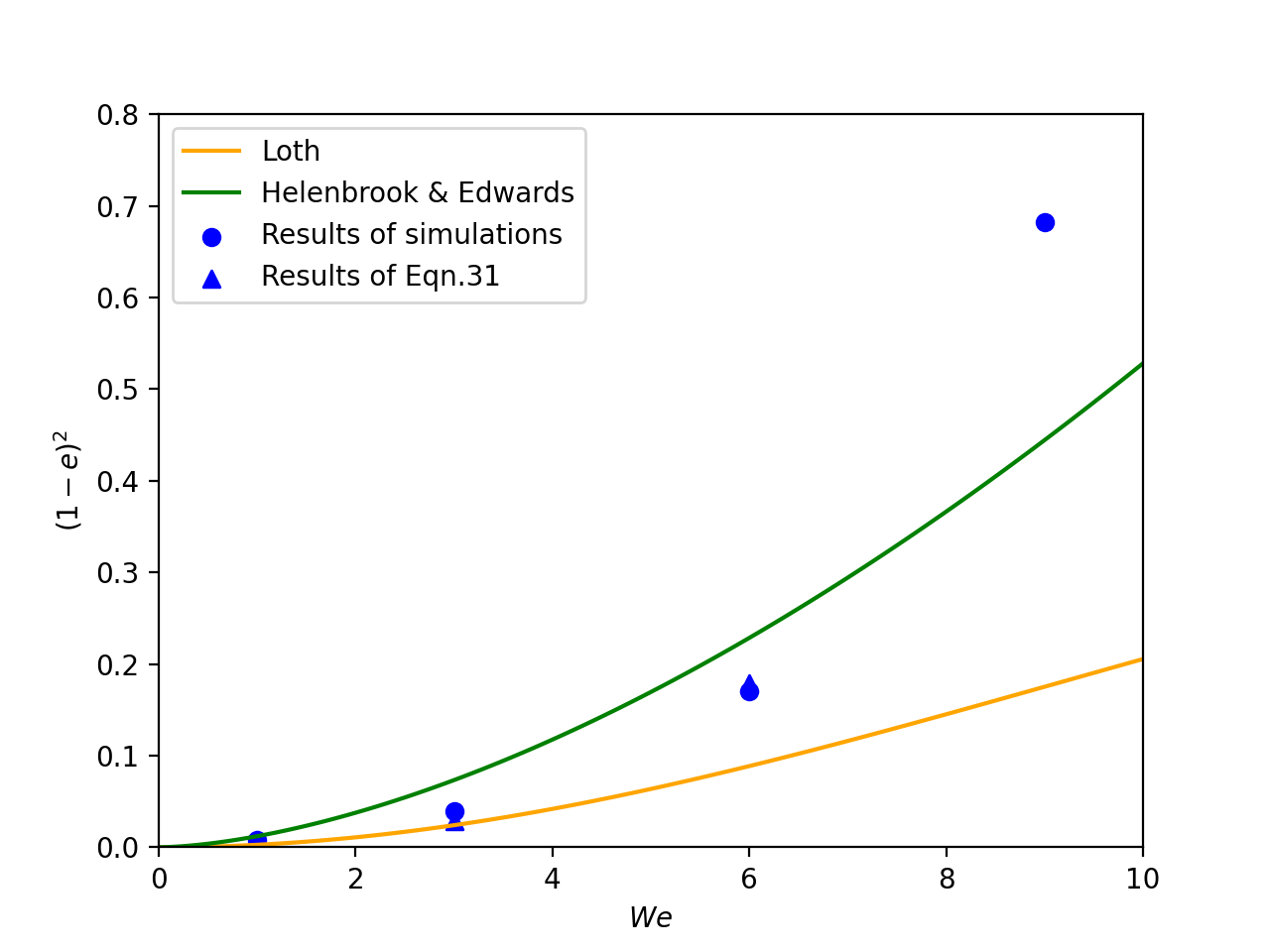}
    \caption{Aspect ratio correlation with $\rho^*/\mu^{*2}$ fixed}
    \label{fixRho}
    \end{subfigure}
    \centering
    \begin{subfigure}{0.49\textwidth}
    \includegraphics[width=1\textwidth]{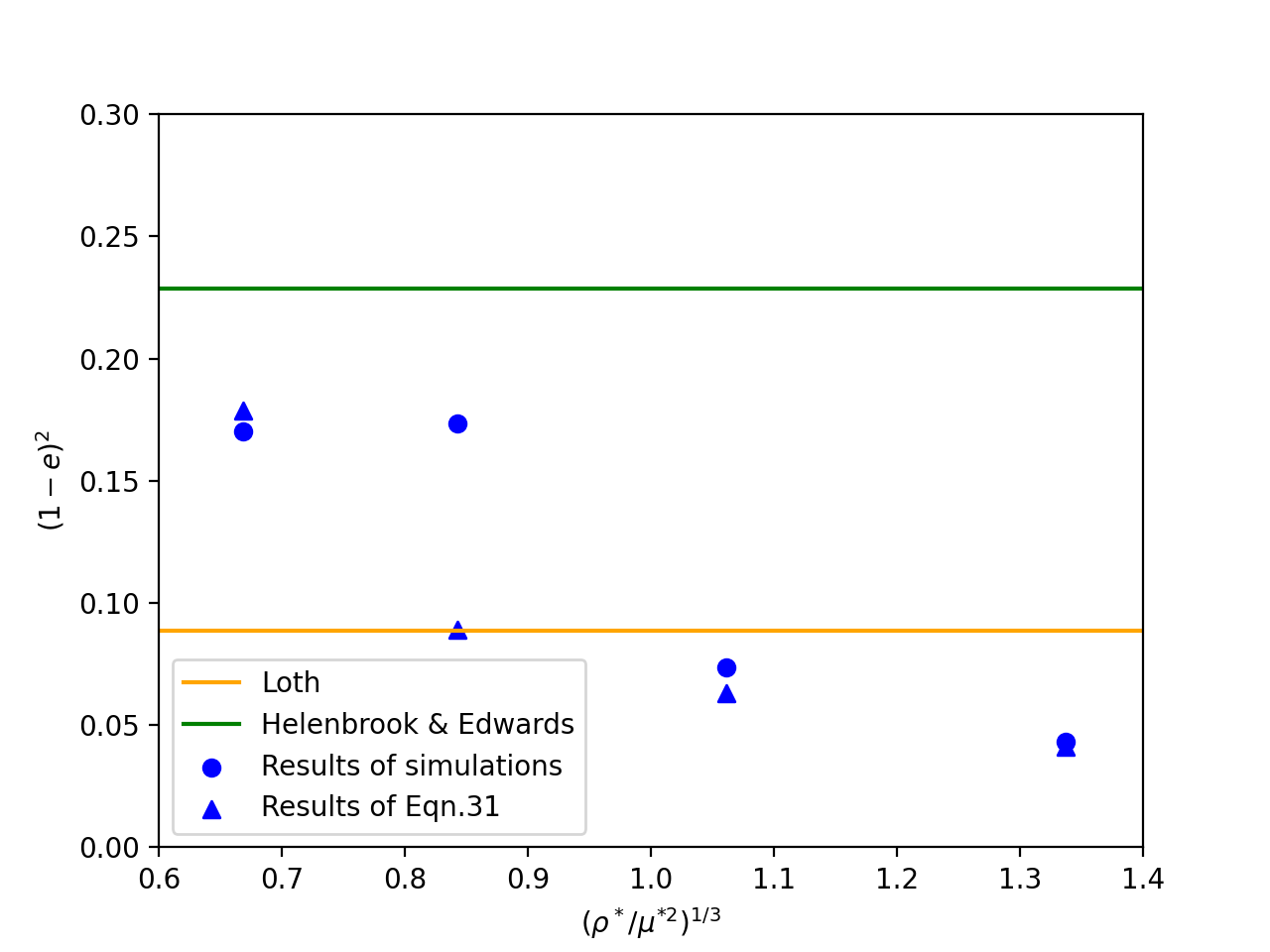}
    \caption{Aspect ratio correlation with $We$ fixed}
    \label{fixWe}
    \end{subfigure}
    \caption{Aspect ratio correlation comparison}
    \label{aspectRatioCorrelation}
\end{figure}

We used data obtained from our numerical simulations by using the in-house code, together with aspect ratio calculated from correlations in Eq.\ref{E_loth} from \cite{loth_quasi-steady_2008} and Eq.\ref{E_helenbrook} from \cite{helenbrook_quasi-steady_2002} to verify Eq.\ref{simplify}. The comparison is shown in Fig.\ref{fixRho}, and $\rho^*/\mu^{*2}$ is fixed at around $0.3$ for the simulations. The data point resulted from our simulations at near $We(1-\rho^*/\mu^{*2})^{1/3}\approx3$, which corresponds to $We=9$, is incorrect. The highly deformed droplets cause the spheroidal assumption to fail, as can be seen in Fig.\ref{drop-we9}. Both correlations in \cite{loth_quasi-steady_2008} and \cite{helenbrook_quasi-steady_2002} are dependent on Weber number only. In \cite{helenbrook_quasi-steady_2002}, there is an aspect ratio correlation dependent on $\rho^*/\mu^{*2}$, but it is only used for prolate droplets. Further, it was developed in cases where liquid-to-gas density ratio is high and viscosity ratio is small. In addition, \citet{helenbrook_quasi-steady_2002} and \citet{feng_drag_2001} have both argued that the effect of $\rho^*/\mu^{*2}$ is minimal when density ratio is small. In our simulations, only droplets with oblate shape were observed, and a significant variation of aspect ratio was found for larger liquid-to-gas density ratio, as can be seen in Fig.\ref{fixWe}, where Weber number was fixed at $We=6$ and viscosity ratio was fixed at around $\mu^*\approx8.18$. When density ratio equals to $20$ and $40$, aspect ratio varies very little. For density ratio equals to $80$ and $160$, aspect ratio decreases significantly. This is reasonable because when ambient pressure is not very high, droplet will copule less to the gas, so it will be less deformed.

\begin{equation}\label{E_loth}
    e=1-0.75\tanh\left(0.07We\right)
\end{equation}
\begin{equation}\label{E_helenbrook}
    e=1-0.11We^{0.82} %+0.013\left(\rho^*/\mu^{*2}\right)^{0.5}Oh^{0.55}We^{1.1}
\end{equation}

\begin{figure}[!ht]
    \centering
    \includegraphics[width=0.6\textwidth]{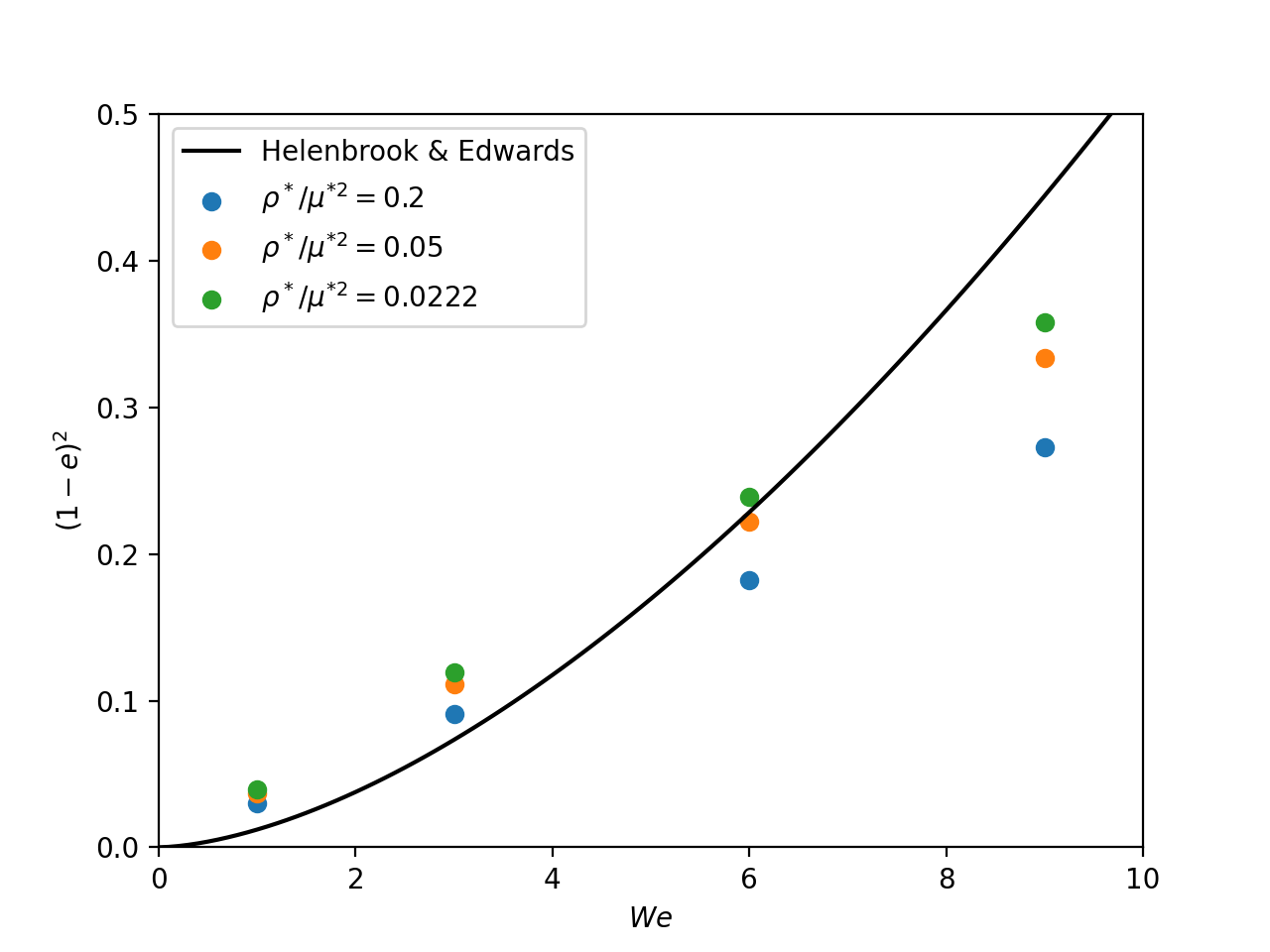}
    \caption{Aspect ratio variation with respect to $\rho^*/\mu^{*2}$ using correlation Eq.\ref{relation_simp_fitted}}
    \label{variation}
\end{figure}

Fig.\ref{aspectRatioCorrelation} indicates Eq.\ref{relation} is able to provide a simple correlation between $We,\ \rho^*/\mu^{*2},\ e$ from a physics perspective. However, since the derivation is not very rigorous, it can only be used to estimate the order of magnitude of $e$. Also, the usage of the correlation is limited to relatively low Weber number to exclude severe deformation situation. Therefore, we further fitted our data to find coefficients for Eq.\ref{relation} to find a more accurate correlation Eq.\ref{relation_fitted}. The results of the fitted correlation is compared in Fig.\ref{aspectRatioCorrelation} as well. However, since the right-hand-side of Eq.\ref{relation} is a concave-up curve with respect to $e$, it will have no solutions when the Weber number is too high, or $\rho^*/\mu^{*2}$ is too small. That is why we do not have the result at $We=9$ of Eq.\ref{relation} in Fig.\ref{fixRho}. An alternative correlation is to fit Eq.\ref{simplify} but it has lower accuracy, and the resulting correlation is Eq.\ref{relation_simp_fitted}. We substituted values of $\rho^*/\mu^{*2}$ used in \cite{helenbrook_quasi-steady_2002} to the correlation to see the variation of aspect ratio, shown in Fig.\ref{variation}. When density ratio is small, aspect ratio won't vary too much. For larger density ratio, the variation of aspect ratio is obvious. The variation also becomes larger with increasing Weber number.

\begin{equation}\label{relation_simp_fitted}
    \left(1-e\right)^2=0.0485We-0.0311We\left(\rho^*/\mu^{*2}\right)^{1/3}
\end{equation}

\begin{equation}
\begin{aligned}\label{relation_fitted}
    -0.0266We\cdot\left(1-0.408\left(\rho^*/\mu^{*2}\right)^{1/3}\right)=\\
    \left(1.14e^{-2/3}-3.11\right)+&\frac{e^{4/3}}{\sqrt{1-e^2}}\ln\left(\left(\frac{1+\sqrt{1-e^2}}{e}\right)^2\right).
\end{aligned}
\end{equation}

\section{Conclusions}
In this work, we investigate the effect of deformation and internal circulation on droplet drag coefficient in high pressure environment. We assume the drag coefficient to be a function of Weber number and liquid-to-gas density, because they reflect droplet deformation and internal circulation, respectively. When calculating droplet drag coefficient, two kinds of estimations are made on the projected frontal area. The first way is to approximate deformed droplet as equal-volume sphere as in many literature do, and the second way is to assume the shape of deformed droplet is spheroid. The second way will fail when the droplet is highly deforemd and about to breakup. The gravity update scheme from \cite{setiya_method_2020,lin_numerical_2022} should force droplet to be steady, but a transient period still exists at the early stage of development. We applied different calculation procedures for the transient acceleration of droplet. We found the acceleration calculated by the time derivative of velocity would be better than the second order time derivative of droplet displacement, the latter oscillating frequently when turning from transient to steady period due to the gravity update scheme. The droplet drag coefficient is found to be larger with increasing Weber number, i.e. stronger deformation, although when the aspect ratio is taken into the definition of the area, this effect becomes very weak. In addition, with decreasing liquid-to-gas density ratio corresponding to stronger internal circulation, the drag coefficient increases. However, it should be noted that deformation and internal circulation are not totally independent of each other, because when droplet is fixed, it still has different aspect ratio with varying density ratio. An anaytical expression is derived which helps explain some of the results. We further explored parameter dependencies of aspect ratio, and found it is correlated with both Weber number and $\rho^*/\mu^{*2}$.

\bibliographystyle{abbrvnat}
\bibliography{1_ref.bib}

\end{document}